\documentclass[%
preprint,
superscriptaddress,
groupedaddress,
 amsmath,amssymb,
]{revtex4-2}

\usepackage{graphicx}
\usepackage{dcolumn}
\usepackage{bm}
\usepackage{epstopdf}
\usepackage{amsmath}
\usepackage{amsfonts}
\usepackage{pifont}
\usepackage[small,bf]{caption}
\usepackage{subfig}
\usepackage{tikz}
\usetikzlibrary{shapes,arrows}
\usetikzlibrary{calc}
\usepackage{pgf}
\usepackage{pgffor}
\usepackage{subfig}
\usepackage{caption}
\usepackage{hyperref}
\usepackage{amssymb}
\usepackage{color,xcolor}
\usepackage{hyperref}
\usepackage[boxruled]{algorithm2e}

\newcommand{\eqnref}[1]{(\ref{#1})}

\newcommand{\trans}{^\mathsf{T}}

\renewcommand{\Re}{\mathbb{R}}
\newcommand{\rom}[1]{\uppercase\expandafter{\romannumeral #1\relax}}

\begin{document}

\title{Reducing transient energy growth in a channel flow using static output feedback control}

\author{Huaijin Yao}
\thanks{yaoxx368@umn.edu}%

\author{Yiyang Sun}%
\author{Talha Mushtaq}%
\author{Maziar S. Hemati}
\affiliation{%
 Aerospace Engineering and Mechanics, University of Minnesota, Minneapolis, MN 55455, USA.
}%

\date{\today}

\begin{abstract}
Transient energy growth of flow perturbations is an important mechanism for laminar-to-turbulent transition that can be mitigated with feedback control.
Linear quadratic optimal control strategies have shown some success in reducing transient energy growth and suppressing transition, but acceptable worst-case performance can be difficult to achieve using sensor-based output feedback control.
In this study, we investigate static output feedback controllers for reducing transient energy growth of flow perturbations within linear and nonlinear simulations of a sub-critical channel flow.
A static output feedback linear quadratic regulator~(SOF-LQR) is designed to reduce the worst-case transient energy growth due to flow perturbations.
The controller directly uses wall-based measurements to optimally regulate the flow with wall-normal blowing and suction from the upper and lower channel walls.
Optimal static output feedback gains are computed using a modified Anderson-Moore algorithm that accelerates the iterative solution of the synthesis problem by leveraging Armijo-type adaptations.
We show that SOF-LQR controllers can reduce the worst-case transient energy growth due to flow perturbations.
Our results also indicate that SOF-LQR controllers exhibit robustness to Reynolds number variations.
Further, direct numerical simulations show that the designed SOF-LQR controllers increase laminar-to-turbulent transition thresholds under streamwise disturbances and delay transition under spanwise disturbances.
The results of this study highlight the advantages of SOF-LQR controllers and create opportunities for realizing improved transition control strategies in the future.

\end{abstract}

\maketitle

\section{Introduction} \label{Sec:intro}
An ability to delay transition to turbulence is of great interest,
owing to the potential for drag reduction and energy savings in numerous engineering systems.
Transient energy growth~(TEG) is an important mechanism for sub-critical
transition in many shear flows~\cite{Schmid2001,Schmid2007}. 
For linearly stable shear flows, small flow perturbations 
can be amplified significantly over short time-horizons~\cite{Trefethen1993,Reddy1993,jovanovic2005}. 
When this TEG is sufficiently large, the flow state can be driven outside
the basin of attraction of the laminar equilibrium,
triggering secondary instabilities that transition the flow to turbulence.
Worst-case analysis is typical in investigations of such phenomena,
since the response leading to the maximum TEG is the one that pushes
the flow state furthest from the laminar equilibrium profile.
The flow perturbation resulting in
the maximum TEG is known as an ``optimal perturbation''~\cite{Butler1992}.
As such, if it is possible to reduce the maximum TEG,
then it may be possible to delay or suppress transition.

Many studies have investigated the possibility of reducing TEG and
suppressing transition by means of feedback control~\cite{Bagheri2009,bagheri2011,joshi1997,Bewley1998,hogberg2003,Martinelli2011,sun2019}.  
Full-state feedback control is usually the first choice for feasibility studies in numerical simulations.
The full-information linear quadratic regulator~(LQR) has been shown to suppress transition in a channel flow using wall blowing and suction actuation~\cite{hogberg2003,Ilak2008,Martinelli2011,sun2019}. 
Worst-case analyses have confirmed that LQR control strategies
reduce the maximum TEG due to linear optimal perturbations.
It was found that another benefit for the LQR controller is that it exhibits robust TEG reduction under off-design Reynolds numbers and wavenumbers~\cite{Kalur2019}.
Despite the successes of full-information LQR control strategies for TEG reduction and transition suppression,
full-state feedback control cannot be realized in practice.
Typically, measurements of the full state are not directly available for feedback; rather,
only measured outputs from a limited set of sensors---typically confined to solid boundaries---can be used for output feedback. 
One commonly used approach for output feedback design is observer-based feedback via the separation principle~\cite{Brogan1991}.
That is, first, an observer is used to estimate the full state of the flow.
Then, this full-state estimate is fed-back to a full-state feedback control law to determine the appropriate actuation.
Indeed, the separation principle of modern linear control theory establishes that observer-based feedback will result in
a stable closed-loop system if the full-state feedback controller is stabilizing and the observer yields stable estimation error dynamics~\cite{Brogan1991}. 
Separation-principle-based designs are appealing owing to the great simplicity of the associated design process:
a controller and estimator can be designed separately, then combined to guarantee closed-loop stability of the linear dynamics. 
For example, the widely used linear quadratic Gaussian~(LQG) controller possesses an observer-based feedback structure that follows the separation principle.
A linear quadratic estimator~(LQE) is combined with a full-information LQR controller to solve
the associated $\mathcal{H}_2$-optimal control problem.
$\mathcal{H}_2$-optimal controllers are designed by solving two independent algebraic Ricatti
equations---one for the optimal observer and one for the optimal controller.
Numerous studies have investigated $\mathcal{H}_2$-optimal controllers and
demonstrated their utility within the context of flow control~\cite{Bewley1998,hogberg2003,bagheri2011}. 

While separation-principle-based designs guarantee closed-loop stability for the linear dynamics,
such designs can result in degraded closed-loop TEG performance~\cite{Hemati2018}.
Indeed, separation-principle-based designs can potentially degrade TEG performance
relative to the uncontrolled system, resulting in adverse consequences for
transition control based on such designs~\cite{hogberg2003,Hemati2018,Yao2018}.
One way of overcoming these limitations is to couple the controller and observer design problems~\cite{Bewley1998}.
Of course, coupling the designs removes the simplicity that is afforded by the separation principle:
acceptable designs require an iterative tuning process that can take substantial effort on the part of the designer.
In principle, the iterative design process can be circumvented by directly seeking
an optimal output feedback controller that minimizes the maximum TEG; however,
such approaches can be computationally intractable~\cite{Whidborne2005,Whidborne2007}.
A potential compromise is introducing additional complexity into the control architecture while maintaining a computationally tractable synthesis problem.
For example, it has been shown that tailored design
kernels and time-varying feedback gains can be
used to achieve substantial TEG reductions~\cite{hapffner2005}.

In this paper, we investigate a simple alternative for TEG reduction within linear and nonlinear simulations of a sub-critical channel flow based on the static output feedback linear quadratic regulator~(SOF-LQR).
SOF-LQR controllers constitute semi-proper control laws, thus satisfying a necessary condition
for overcoming the TEG performance limitations of observer-based feedback~\cite{Hemati2018,Whidborne2007}.
Specifically, we solve the standard LQR problem, but with an additional constraint 
that the control law has a static output feedback control structure.
The SOF-LQR controller gain in this study maps the wall-based measurements at the channel walls
directly to the control input, which is taken to be the rate of change of wall-normal velocity at upper and lower walls. 
In order to expedite the computations involved in determining the optimal SOF-LQR gains, we introduce an accelerated
gradient method based on the Anderson-Moore algorithm~\cite{Anderson1971}.
Controllers are designed using various combinations of wall-based shear-stress and pressure sensors.
The resulting SOF-LQR controllers reduce the worst-case TEG and exhibit robustness to Reynolds number and wavenumber uncertainties. Furthermore, the non-linear performance of SOF-LQR control is investigated using nonlinear direct numerical simulations~(DNS).
The DNS results show that the proposed SOF-LQR control delays transition under spanwise disturbances and increases transition thresholds under streamwise disturbances. 
As such, SOF-LQR stands as a viable candidate for transition delay and suppression for future studies.

The remainder of the paper is organized as follows. Full-information LQR and SOF-LQR control strategies for TEG reduction are discussed in Section~\ref{Sec:feedbacks}.
We also introduce an accelerated Anderson-Moore algorithm for designing SOF-LQR controllers in this section.
In section~\ref{sec:model}, we present the linearized channel flow model and give the setup for the direct numerical simulation~(DNS). The results are sequenced by disturbance type (i.e.,~spanwise and streamwise) in section~\ref{sec:results}.
Performance and robustness of SOF-LQR control for TEG reduction are investigated using linear simulations.
Transition suppression and delay scenarios and mechanisms are investigated using nonlinear simulations.
Comparisons are made with the uncontrolled flow and full-information LQR control.
Finally, conclusions are presented in section~\ref{sec:conclusions}.

\section{Transient energy growth and controller synthesis}  \label{Sec:feedbacks}
\subsection{Transient energy growth}
Consider the state-space representation of the linearized Navier-Stokes equations about a laminar equilibrium solution, 
\begin{equation}
\begin{aligned}
\dot{X}(t) &=AX(t)+BU(t) \\   \label{eq:op_system}
Y(t) &= CX(t)  
\end{aligned}
\end{equation}
where $X\in\Re^n$ is the state vector, $U\in\Re^m$ is the input vector, 
$Y\in\Re^p$ is the output vector, and $t\in\Re$ is time. 
For an initial flow perturbation $X(t_0)=X_0$, the system response is given in terms of the matrix exponential $X(t)={e}^{A(t-t_0)}X_0$, where $A$ represents the system dynamics matrix.
The associated perturbation kinetic energy is given as,
\begin{equation}
E(t)=X\trans(t)QX(t),  \label{eqn:Q_define}
\end{equation}
where $Q=Q\trans > 0$. 
Further, the maximum TEG is defined as,
\begin{equation}
G=\max _{t \geq t_0}\max _{E(t_0)  \neq 0} \frac{E(t)}{E(t_0)}, 
\end{equation}
which results from a so-called \emph{worst-case} or \emph{optimal perturbation}~\cite{butlerPOF1992}.
Certain perturbations will result in non-trivial TEG whenever $G>1$.

\subsection{Full-information feedback control synthesis}
Feedback controllers have been shown to reduce TEG in various shear flows.
In particular, the linear quadratic regulator~(LQR) is a well-known design technique 
that has been proven to be successful at reducing TEG in previous flow control studies~\cite{Ilak2008,Martinelli2011,sun2019}.
LQR synthesis is based on solving,
\begin{equation}
\min_{U(t)} J = \int _0 ^\infty \left[X\trans(t)QX(t)+U\trans(t)RU(t)\right]dt \label{eqn:objective_function}
\end{equation}
subject to the linear dynamic constraint
\begin{equation}
\dot{X}(t) = AX(t)+BU(t)
\end{equation}
where $R>0$. 
The resulting LQR controller is a full-state feedback law of the form $U(t)=KX(t)$, where $K\in\Re^{m\times n}$ is determined from the solution of an algebraic Riccati equation~\cite{Brogan1991}.
LQR controllers are particularly appealing because they can be designed to reduce TEG, while demonstrating robustness to parametric and modeling uncertainties. 
However, outside of numerical simulations, standard full-state feedback LQR controllers are typically not practically viable for flow control; standard LQR control requires knowledge of the full state of the flow, 
which is usually not directly available for feedback in practice.

When full-state feedback is not a viable option, an observer (i.e.,~state estimator)
is usually designed to estimate the current state of the flow from available sensor measurements $Y(t)$.
The separation principle is often invoked to simplify the design process;
however, doing so can degrade the resulting closed-loop TEG performance~\cite{Hemati2018,hogberg2003,hapffner2005,Yao2018,Yao2019}. 
This performance degradation arises due to unaccounted adverse interactions that can
arise between the fluid dynamics and the control system dynamics~\cite{Hemati2018}.
The control system dynamics can be designed to overcome this performance degradation,
but proposed strategies introduce complexity in the design procedure and/or in the control law~\cite{hogberg2003,Whidborne2007}.
To address the TEG reduction problem with sensor-based feedback, we introduce an alternative static output feedback control strategy that provides a simple alternative to overcome the performance limitations of the separation principle and associated observer-based designs.

\subsection{Static output feedback control synthesis}
The proposed SOF-LQR control strategy for TEG reduction is based on
solving the standard LQR problem,
but now with an additional constraint that the resulting feedback law have static output feedback~(SOF) control structure,
\begin{equation}
U(t)=FY(t), \label{eqn:SOF_controller}
\end{equation}
where $F \in \mathbb{R}^{m \times p}$ is the SOF gain matrix.
The benefit of the SOF control structure is that the input is determined directly from the measured output,
removing the need for state estimation.
We note that the SOF control structure is akin to the so-called ``opposition control'' that has been employed and studied within the context of turbulent drag reduction~\cite{choi1994,luhar2014}.
Furthermore, the SOF structure constitutes a semi-proper controller, thus satisfying a
necessary condition for eliminating TEG~\cite{Hemati2018,Whidborne2007}.
As such, we expect the SOF-LQR controller to improve performance worst-case TEG relative to strictly proper control structures,
such as LQG controllers and other observer-based feedback controllers designed via the separation principle.

The closed-loop dynamics under SOF-LQR control are of the form,
\begin{equation}
\dot{X}(t)=(A+BFC)X(t).
\end{equation}
Thus, the standard LQR objective function in~\eqref{eqn:objective_function} can be rewritten to conform to the SOF structure in~\eqref{eqn:SOF_controller} as,
\begin{equation}
J=\int _0 ^ \infty X\trans(t)[Q+(FC)\trans R(FC)]X(t)dt.  \label{eqn:sof_obj}
\end{equation}
The solution of this SOF-LQR problem can be calculated iteratively using Anderson-Moore methods~\cite{Anderson1971}.
To do so, define the set of all stabilizing SOF controllers $D_s = \{F \in \mathbb{R}^{m \times p} | \textit{Re} \{ \lambda(A+BFC) \}<0 \}$, where $\lambda(\cdot)$ denotes the set of all eigenvalues of $(\cdot)$.
Then, re-write the SOF-LQR design problem as~\cite{Rautert1997,Syrmos1997},
\begin{equation}
 \min_F \; \; J(F)=\textrm{trace}[S(F)X_E] \;\; \text{subject to} \; \;  F\in D_s,   \label{eqn:design_problem}
\end{equation} 
where $X_E=\mathbb{E}\{X(0)X(0)\trans\}$ and $S(F)$ is a solution to the algebraic Ricatti equation,
\begin{equation}
S(F)[A+BFC]+[A+BFC]\trans S(F)+C\trans F\trans RFC+Q = 0. \label{eqn:solve_for_S}
\end{equation}
The gradient of the cost function~$J$ with respect to the SOF control gain~$F$ can be expressed as
\begin{equation}
\frac{\partial J}{\partial F} = 2[B\trans S(F)H(F)C\trans+RFCH(F)C\trans], \label{eqn:gradient_terminate}  
\end{equation}
where $H(F)$ is the solution to the Lyapunov equation
\begin{equation}
H(F)[A+BFC]\trans+[A+BFC]H(F)+X_E=0.    \label{eqn:solve_for_H}
\end{equation}
Then,
the optimal SOF gain $F^* \in D_s$ will be a minimizer of \eqref{eqn:design_problem} and so must satisfy a zero gradient condition, which reduces to
\begin{equation}
[B\trans S(F^*)H(F^*)C\trans +RF^*CH(F^*)C\trans]=0.
\end{equation}
After some further manipulation, 
we find that a necessary condition for optimality is 
\begin{equation}F = -R^{-1}[B\trans S(F)H(F)C\trans][CH(F)C\trans]^{-1},\end{equation}
yielding a search direction to use in the Anderson-Moore method,
\begin{equation}
T = -F-R^{-1}[B\trans S(F)H(F)C\trans ][CH(F)C\trans ]^{-1}.
\label{eqn:searchdir}
\end{equation}

Although the expressions above are sufficient for implementing
the Anderson-Moore method, such methods tend to require 
a significantly large number of iterations.
For high-dimensional fluid flows, 
each iteration can require a 
significant computational demand on the order of $\mathcal{O}(n^3)$,
and so it is 
desirable to reduce the total number of iterations 
through an accelerated technique.
The step-size $\xi$ along the gradient direction must be chosen with care in order to balance precision with the 
total number of iterations.
An inappropriate choice of $\xi$ will
lead to slow convergence.
In order to overcome this challenge, we formulate 
an accelerated Anderson-Moore algorithm that incorporates 
Armijo-type adaptations. 
Instead of using a fixed step-size $\xi$, 
we instead use an Armijo-rule~\cite{nocedal2006} to 
adaptively update the step-size to
achieve a better balance between precision 
and iteration count.
The method we propose and use in this study is summarized as Algorithm~1. 

\begin{center}
\begin{minipage}[t]{0.85\textwidth}
\begin{algorithm}[H] \label{Algorithm1}

\textbf{step 0:} Set i = 0; initialize $F_i=F_0$ to be any $F_0\in D_s$.  Set $0<\xi<1$, $0<\sigma<1/2$, and $\delta>0$. 

\textbf{step 1:} Solve \eqref{eqn:solve_for_S} for $S(F_i)$.

\textbf{step 2:} Solve \eqref{eqn:solve_for_H} for $H(F_i)$.

\textbf{step 3:} Use~\eqref{eqn:searchdir} to find the smallest integer $\gamma_1 \geq 1$ such that ${F_i+\xi ^{\gamma_1}T_i} \in D_s$.

\textbf{step 4:} Find the smallest integer $\gamma_M \geq \gamma_1$ such that 
\begin{equation*}J(F_i+\xi ^{\gamma_M}T_i) \leq J(F_i)+\sigma \xi ^{\gamma_M} \mathrm{trace}(\frac{\partial J}{\partial F}\trans T_i).\end{equation*}

\textbf{step 5:} Find integer $\ell\in\{\gamma_1,\dots,\gamma_M\}$ such that 
\begin{equation*}
J(F_i+\xi ^{\ell}T_i) =  \min J(F_i+\xi ^{j}T_i),\text{ where } 
j \in \{ \gamma_1,\cdots, \gamma_M \}
\end{equation*}

\textbf{step 6:} Set $F_{i+1} = F_i+\xi ^\ell T_i$, $i = i+1$

\textbf{step 7:} Check $\Vert \frac{\partial J}{\partial F} \Vert _2 \leq \delta$. If true, stop. Otherwise, go to \textbf{step 1}.\\
\caption{\textbf{\textit{Anderson-Moore algorithm with Armijo-type adaptation}}}
\end{algorithm}
\end{minipage}
\end{center}

Note that all Anderson-Moore methods require initialization with 
a stabilizing SOF gain $F_0 \in D_s$~\cite{Toivonen1985}.
For an asymptotically stable system, setting $F_0=0$ is a valid choice.
In the present study, actuator dynamics are modeled via an integral term, and so the 
resulting linear system model will not be asymptotically stable.
Thus, we first determine a stabilizing static output feedback gain $F_0$ using the 
iterative linear matrix inequality (ILMI)
method proposed in \cite{Cao1998a}, then proceed to compute the optimal SOF-LQR controller using the accelerated Anderson-Moore algorithm in Algorithm~1. 
Additional details on the ILMI method used for initialization are presented in Appendix~\ref{sec:app1}. 
The computational complexity of Algorithm~1 is detailed in Appendix~\ref{sec:app2}.

\section{Channel flow model}
\label{sec:model}
The proposed SOF-LQR controller will be evaluated using both linear analysis and nonlinear direct numerical simulations in section \ref{sec:results}. In this section, we present the linearized channel flow model used for controller design and provide details of the direct numerical simulations.
\subsection{Linearized channel flow system} \label{sec:fluid_model}
\begin{figure}
\centering
\includegraphics[width=0.6\textwidth]{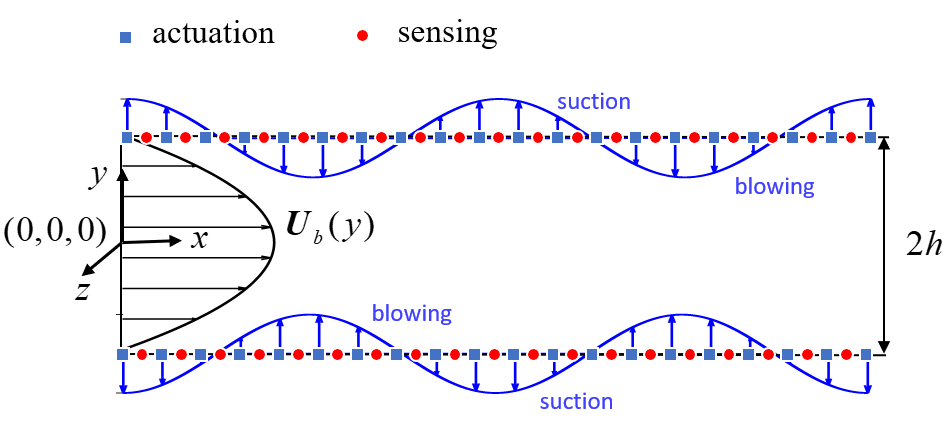}
\caption{Illustration of wall-based sensing and actuation in channel flow.} \label{fig:channel_flow}
\end{figure}

Consider the pressure-driven flow between two infinite parallel
walls separated by a distance $2h$, shown in Figure~\ref{fig:channel_flow}.
Here, $x$, $y$, and $z$ represent the streamwise, wall-normal, and spanwise directions, respectively. The laminar equilibrium solution of this plane Poiseuille flow is a parabolic profile in the form of $\boldsymbol{U}_b({y})=\bar{u}_c(1-{y}^2/h^2)$, 
where $\bar{u}_c$ is the centerline velocity of the base flow.
The incompressible Navier-Stokes and continuity equations for small perturbations are linearized about this laminar profile. Transforming into velocity-vorticity form and Fourier transforming in the streamwise and spanwise directions, yields the Orr-Sommerfeld and Squire equations for the linear perturbation dynamics~\cite{Schmid2001}:
\begin{equation}
\begin{aligned}
   (\mathcal{D}^2-k^2) \dot{\tilde{{v}}} &= \left[- i \alpha \boldsymbol{U}_b( \mathcal{D}^2-k^2)+i\alpha \boldsymbol{U}_b''+ \frac{1}{Re} (\mathcal{D}^2-k^2)^2\right]\tilde{{v}}  \\
    \dot{\tilde{{\eta}}} &=\left[- i\alpha \boldsymbol{U}_b+\frac{1}{Re}( \mathcal{D}^2-k^2)\right]\tilde{{\eta}} -i\beta \boldsymbol{U}_b'\tilde{{v}},
\end{aligned} \label{eqn:v_v}
\end{equation} 
where ${v}$ is the wall-normal velocity,
${\eta}$ is the wall-normal vorticity, $\tilde{(\cdot)}$ denotes the Fourier amplitude of the associated variable,
$\mathcal{D}$ represents differentiation with respect to ${y}$, and  $k^2 = \alpha^2+\beta^2$, where $\alpha$ and $\beta$ are wavenumbers of the streamwise and
spanwise Fourier modes, respectively.
$Re$ denotes the Reynolds number based on the channel half-height.
No-slip boundary conditions are imposed at the solid walls for uncontrolled flow, i.e., $\tilde{{v}}=\mathcal{D}\tilde{{v}}=\tilde{{\eta}}=0$.
Next, $\tilde{{v}}$ and $\tilde{{\eta}}$ can be approximated
by a finite Chebyshev series expansion along the wall-normal direction
with $N$ discrete collocation points.
For this uncontrolled channel flow,
the state variable is $X_u = (a_{{v}0},\dots,a_{{v}N},a_{{\eta} 0},\dots,a_{{\eta} N})\trans $,
where $a_i$ are the Chebyshev polynomial coefficients.

For flow control, actuation is achieved using wall-normal
velocity at the upper and lower walls via
wall-transpiration boundary conditions denoted as
$\tilde{{v}}|_{+h}$ and $\tilde{{v}}|_{-h}$.
The control input is taken to be the rate of wall-normal blowing and suction at each wall.
Thus, the input vector in~\eqref{eq:op_system} is ${U}=\frac{\partial}{\partial t}[\hat { v}|_{+h}, \hat { v}|_{-h}]^T$, and
the associated system state vector consists of
the Chebyshev polynomial coefficients and the actuator states as
\begin{equation}
    X = (a_{{v}0},\dots,a_{{v}N},a_{{\eta} 0},\dots,a_{{\eta} N},\tilde{{v}}|_{+h}, \tilde{{v}}|_{-h})\trans.
\end{equation}
We consider several wall-based sensors in this study, each of which can be represented in Fourier space.
Shear-stress measurements are given by
\begin{equation}
  \begin{aligned}
  \tilde{\tau}_{{x}} \vert _{{y}=\pm h}  &=\frac{1}{Re}\left(\frac{\partial {\tilde{u}}}{\partial {y}}\right), \\
  \tilde{\tau}_{{z}}\vert _{{y}=\pm h}  &=\frac{1}{Re}\left(\frac{\partial {\tilde{w}}}{\partial {y}}\right),
  \end{aligned} \label{eqn:sensor_dudy}
\end{equation}
where $\tilde{u}$ and $\tilde{w}$ correspond to Fourier coefficients of streamwise and spanwise velocity components, respectively. 
We also consider spatial derivatives of the shear-stress with respect to the wall-normal direction, 
\begin{equation}
    \begin{aligned}
     \left.\frac {\partial \tilde{\tau}_{{x}}}{\partial y} \right\vert_{{y}=\pm h} &=\frac{1}{Re}\left(\frac{\partial^2 {\tilde{u}}}{\partial {y}^2}\right), \\
     \left.\frac {\partial \tilde{\tau}_{{z}}}{\partial y}\right \vert _{{y}=\pm h} &=\frac{1}{Re}\left(\frac{\partial^2 {\tilde{w}}}{\partial {y}^2}\right).
    \end{aligned} \label{eqn:sensor_du2dy}
\end{equation}
Lastly, we consider the pressure measurements at the upper and lower channel walls.
 The Fourier coefficient for pressure can be expressed in terms of the wall-normal velocity and vorticity using the $x$- and $z$-momentum equations.  
    \begin{equation}
        \left.\tilde{p}\right|_{y=\pm h}=\frac{1}{\alpha^2+\beta^2}\frac{1}{Re}\left(\frac{\partial ^3\tilde{v}}{\partial y^3}+\frac{\alpha-\beta}{\alpha+\beta}\frac{\partial^2\tilde{\eta}}{\partial y^2}\right)+\frac{i}{\alpha+\beta}\tilde{v}\frac{\partial \boldsymbol{U}_b}{\partial y}. \label{eqn:sensor_p}
    \end{equation}
Further details about the model formulation can be found in~\cite{McKernanJ2006}. 

For sensor-based output feedback control, we investigate three different sensor combinations among the quantities reported in equations \eqref{eqn:sensor_dudy}-\eqnref{eqn:sensor_p}. Specific configurations are listed in Table~\ref{tab:sensors}, and will be referenced accordingly in the remainder of this paper.

\begin{table}
\caption{Wall-based sensor configurations}
\label{tab:sensors}
\begin{tabular}{ |p{6.5cm}|p{6.5cm}| }
\hline
Configuration & Sensors \\
\hline
s &  $\tilde{\tau}_{{x}},\tilde{\tau}_{{z}}$  \\\hline
sp & $\tilde{\tau}_{{x}},\tilde{\tau}_{{z}},\tilde{p} $ \\\hline
ssdp &  $\tilde{\tau}_{{x}},\tilde{\tau}_{{z}},  \frac {\partial \tilde{\tau}_{{x}}}{\partial y},\frac {\partial \tilde{\tau}_{{z}}}{\partial y} ,\tilde{p}$ \\\hline
\end{tabular} 
\end{table}

\subsection{Direct numerical simulation setup} \label{sect:dns_setup}

 Three-dimensional direct numerical simulations~(DNS) of plane Poiseuille flow are performed to analyze the nonlinear performance of the designed controllers. 
 The incompressible Navier-Stokes equations are solved using a modified version of the spectral code \textit{Channelflow} \citep{channelflow,sun2019}. 
The flow response to optimal disturbances for a given controller is simulated. 
A second-order semi-implicit Crank-Nicolson Runge–Kutta temporal scheme is used. 
Spanwise and streamwise disturbance scenarios are considered. 
The kinetic energy density of the initial optimal perturbation is denoted by $E_0$. For each wavenumber pair, several amplitudes of $E_0$ are considered to demonstrate the role of the nonlinearity in the laminar-to-turbulent transition. 
A random disturbance with perturbation kinetic energy density of $1\%$ of $E_0$ is superposed with the optimal disturbance profile to ensure that a laminar-to-turbulent transition can be initiated~\cite{Reddy:JFM98}. 
A three-dimensional computational domain is used in order to resolve the complete transition process.
We use a rectangular computational domain of size $8\pi h \times 2h \times 2\pi h$ in $x$-, $y$- and $z$-directions, respectively. 
To discretize the flow field, $N=101$ Chebyshev points are specified in the $y$-direction, and $128\times64$ points are uniformly spaced along $x$- and $z$-directions, respectively. 
For both baseline and controlled flows, grid resolution studies with doubled grids in each direction have been performed to ensure the accuracy of results.  

\section{Results} \label{sec:results}
The utility of SOF-LQR control for TEG reduction will first be evaluated using linear simulations.
DNS will then be performed to evaluate the nonlinear performance of SOF-LQR control for transition suppression and delay.
We design SOF-LQR controllers for a sub-critical Reynolds number $Re=3000$, then evaluate the worst-case performance associated with spanwise $(\alpha,\beta)=(0,2)$ and streamwise $(\alpha,\beta)=(1,0)$ linear optimal perturbations. 
As shown in section~\ref{sec:fluid_model}, the dynamical systems of the uncontrolled and controlled flows are different;  
thus, the optimal disturbances are calculated independently for each system to ensure a fair comparison based on the largest TEG under each setting. 
In the nonlinear DNS, as described in section~\ref{sect:dns_setup}, we adopt the optimal disturbances calculated for the linearized channel flow and add a small random perturbation to trigger the laminar-to-turbulent transition. We investigate three different sensor combinations for feedback control. These are listed in Table~\ref{tab:sensors}.

\subsection{Spanwise disturbances}
\subsubsection{Linear analysis}
\begin{figure}[h!]
\centering
\includegraphics[width=0.475\textwidth]{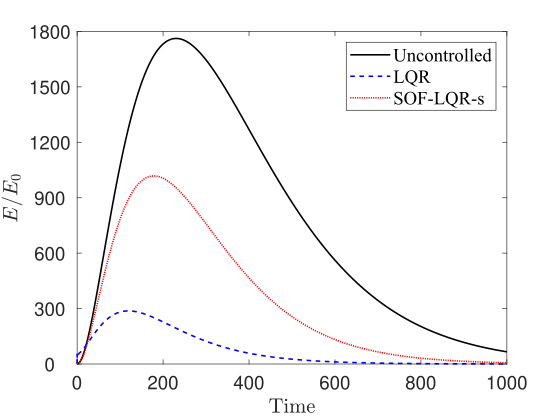}
\caption{Linear worst-case response to spanwise optimal perturbations with $(\alpha,\beta)=(0,2)$ and $Re=3000$.}
\label{fig:a0b2_TEG}
\end{figure}

For spanwise disturbances, we investigate SOF-LQR control using sensor configuration ``s'' in Table~\ref{tab:sensors}, which consists solely of shear-stress measurements $\tilde{\tau}_{{x}}$ and $\tilde{\tau}_{{z}}$.
The linear worst-case response for SOF-LQR-s control with $(\alpha,\beta) = (0,2)$ and $Re=3000$ is reported in Figure~\ref{fig:a0b2_TEG}~(red dotted line).
The SOF-LQR-s controller reduces the maximum TEG~($G$) by approximately 50\% relative to the uncontrolled flow~(black solid line). 
The worst-case response for the full-information LQR controlled system ~(blue dashed line) yields a larger TEG reduction, achieving approximately $80\%$ reduction relative to the uncontrolled flow.
This result is not surprising since the SOF-LQR-s design operates based on considerably less information than the full-information case.
As we will see momentarily, this reliance on limited information actually improves the robustness of the flow control strategy to parameter and modeling uncertainties.
We will discuss physical mechanisms surrounding the TEG reduction within the context of nonlinear simulations after we report results on flow control robustness.

Thus far, we have considered control performance at a fixed Reynolds number of $Re=3000$ and a fixed spanwise disturbance $(\alpha,\beta)=(0,2)$. 
In practice, the fluid flow could experience a mix of disturbances, and the underlying parameters may not be precisely known. 
As such, we conduct additional linear simulations to better characterize and understand the robustness of SOF-LQR-s performance to other spanwise disturbances and Reynolds numbers.
In these studies, we will design the controller under the assumption of $Re=3000$ and $(\alpha,\beta)=(0,2)$, but then analyze the response to ``off-design'' disturbances and Reynolds numbers.
We again consider
the flow response to an optimal perturbation that results in
maximum TEG~($G$) for the associated closed-loop system.
\begin{figure}
\subfloat[Controllers designed for $(\alpha,\beta)=(0,2)$ ]{\label{fig:robustness_a0b2}
\includegraphics[width=0.4\textwidth]{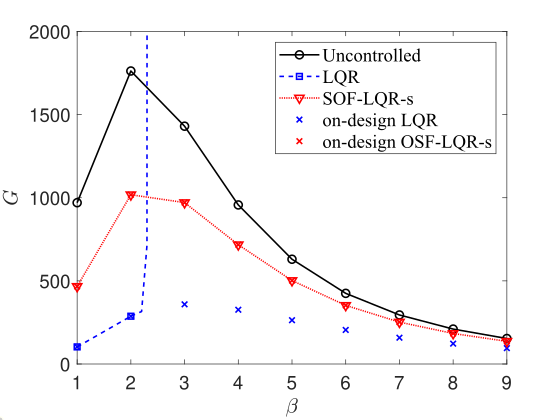}
} 
\hspace{1pt}
\subfloat[Controllers designed for $(\alpha,\beta)=(0,5)$]{\label{fig:robustness_a0b5}
\includegraphics[width=0.4\textwidth]{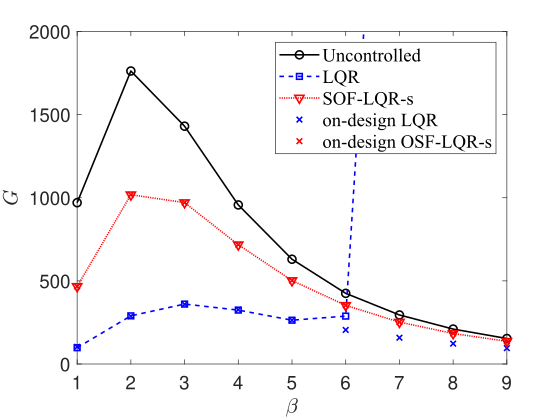}
}
\caption{Robustness of LQR and SOF-LQR controllers to off-design spanwise perturbations at $Re=3000$.} \label{fig:streamwise_robustness}
\end{figure}

We first consider the controller robustness to perturbation-wavenumber variations. In figure~\ref{fig:robustness_a0b2}, the LQR and SOF-LQR-s controllers are designed for wavenumber pair $(\alpha,\beta)=(0,2)$ at $Re=3000$, and the controller is applied at off-design wavenumber conditions of $\alpha=0$ and $\beta=[1,9]$ at $Re=3000$. 
In conducting this study, we observe a remarkable finding:
the full-information LQR controller results in
an unstable closed-loop response for $\beta\ge2.5$,
whereas the SOF-LQR-s maintains comparable performance to
the on-design responses as marked by the black crosses.
It seems that under spanwise disturbances, the same richness of information that results in superior on-design TEG performance is deleterious
to robust performance.
In full-information control, modeling uncertainties
contaminate the control through many more channels
than is possible for the static output feedback case.
The fact that the SOF-LQR-s controller leverages only limited information about shear-stress at the walls can limit TEG performance but also lends itself to robustness against modeling uncertainties. 
We extend this analysis further by considering a controller designed at the on-design condition of $(\alpha,\beta)=(0,5)$, then applying it over the same set of off-design wavenumbers (see figure~\ref{fig:robustness_a0b5}).
Interestingly, the full-information LQR controller exhibits robust performance when applied to off-design conditions with spanwise wavenumber less than the on-design value (i.e.,~$\beta\le5$).
However, when applied to mitigate spanwise disturbances at off-design wavenumbers larger than the on-design wavenumber (i.e.,~$\beta>5$), even the closed-loop stability of the same full-information LQR controllers is lost. 

\begin{figure}
\includegraphics[width=0.475\textwidth]{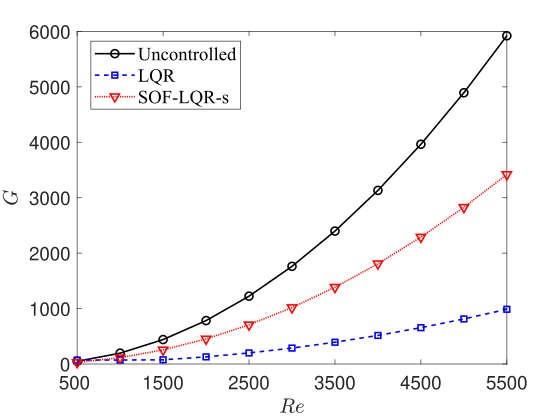}
\caption{Robustness of LQR and SOF-LQR-s controllers to off-design Reynolds numbers. Both controllers are designed for $Re=3000$, but robustly reduce TEG from spanwise optimal perturbations over a range of off-design Reynolds numbers.} 
\label{fig:spanwise_robustness_to_Re}
\end{figure}

Next, we consider controller performance at off-design Reynolds numbers, keeping the disturbance wavenumber pair at the ``on-design'' condition.
Specifically, the SOF-LQR-s and full-information LQR controllers
are designed for $Re=3000$ with $(\alpha,\beta)=(0,2)$ for spanwise disturbance.
Then the controllers are applied at ``off-design'' Reynolds numbers in the sub-critical range of $Re=[500,5500]$ (see Figure~\ref{fig:spanwise_robustness_to_Re}).
The lower limit is based on the observations of
turbulence onset at $Re=500$~\cite{orszag_kells_1980};
the upper limit is chosen to be slightly less than the
critical Reynolds number for linear instability, $Re_c=5772$~\cite{orszag_1971}.
The results in Figure~\ref{fig:spanwise_robustness_to_Re} indicate that both controllers are able to reduce the maximum TEG at off-design Reynolds numbers.
The relative reduction is comparably smaller at lower $Re$---owing to the low degree of TEG in the uncontrolled flow---and becomes more pronounced at larger $Re$.
These results demonstrate robust control performance to Reynolds number variations in the context of spanwise disturbances. 
\subsubsection{Nonlinear direct numerical simulation}
\begin{figure}
    \centering
    \includegraphics[width=0.6\textwidth]{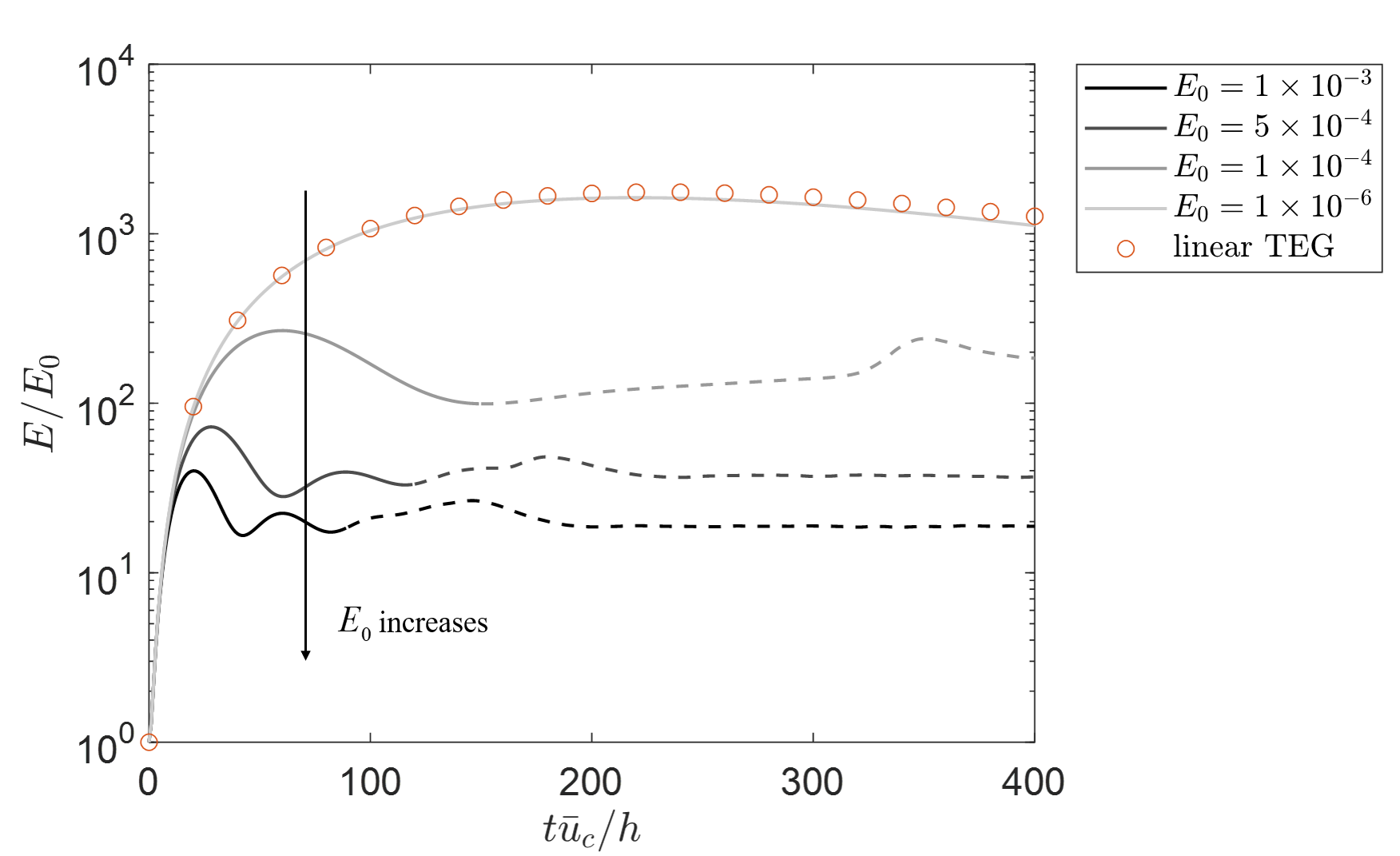}
    \caption{($\alpha,\beta$) = (0,2) transient energy growth of optimal disturbance in uncontrolled flow with various amplitudes of $E_0$.}
    \label{fig:a0b2_base}
\end{figure}

Nonlinear direct numerical simulations~(DNS) are performed under the optimal disturbance associated with Reynolds number $Re=3000$ and wavenumber pair $(\alpha,\beta)=(0,2)$.
We start the nonlinear simulation by initializing the flow field with the base flow and the optimal disturbance resulting from the linear analysis. The amplitude of the perturbation's kinetic energy density $E_0$ is implemented over the range from $1\times 10^{-6}$ to $1 \times 10^{-3}$. The smallest amplitude of the kinetic energy density of disturbance $E_0=1\times 10 ^{-6}$ considered overlaps the linear result, as shown in figure~\ref{fig:a0b2_base}.
To trigger laminar-to-turbulent transition, a random perturbation of $1\%$ of $E_0$ is added to the optimal disturbance.
The nonlinear effects are illustrated by reporting the normalized perturbation energy $E/E_0$. 
For the spanwise disturbance, as shown in figure~\ref{fig:a0b2_base}, the uncontrolled flow undergoes a laminar-to-turbulent transition for kinetic energy density at and above the threshold of $E_0 = 1 \times 10^{-4}$.

\begin{figure}
    \includegraphics[width=1\textwidth]{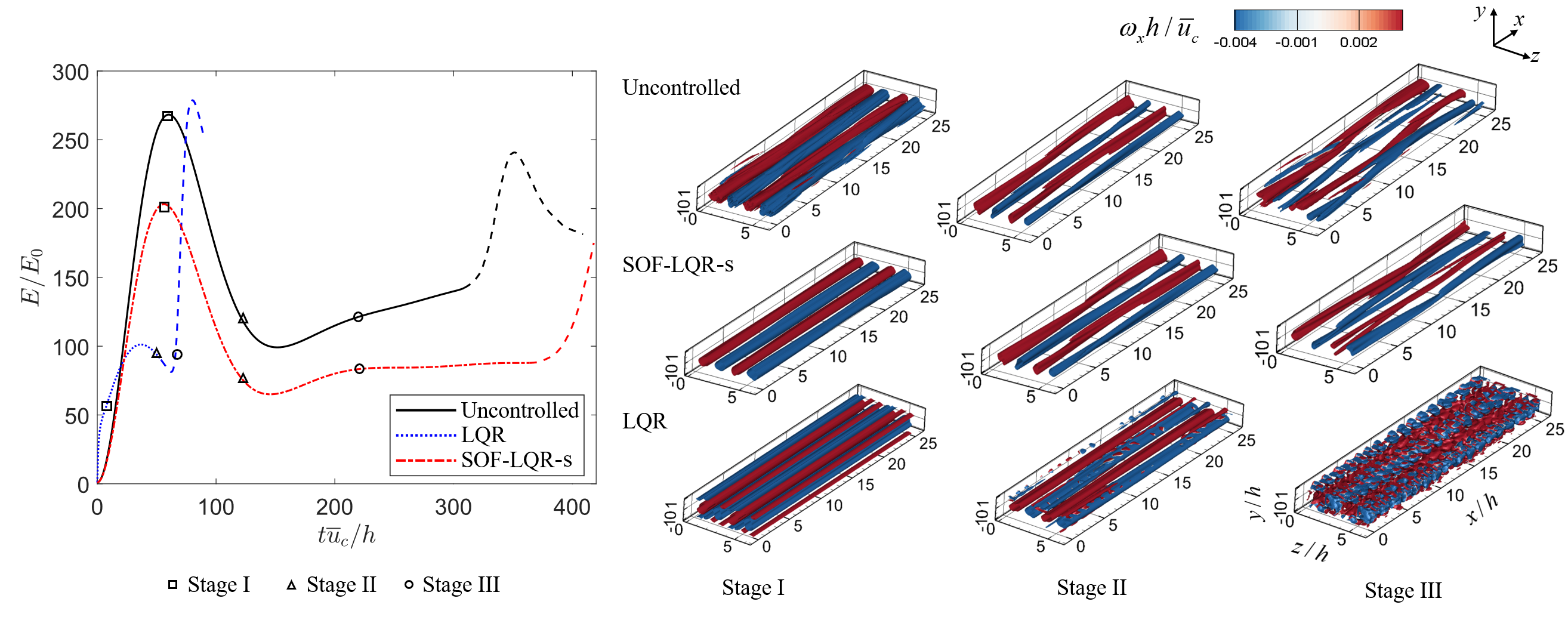}
    \caption{Transient energy growth of uncontrolled and controlled flows due to linear optimal perturbations of $E_0 =1\times 10^{-4}$. Inserts correspond to iso-surfaces of $\mathcal{Q}$-criterion~\citep{Hunt:88}~($\mathcal{Q}(h/\bar{u}_c)^2=4\times 10^{-4}$) colored by streamwise vorticity $\omega_x h/\bar{u}_c$.}
    \label{fig:a0b2_base_flow fields}
\end{figure}

In figure~\ref{fig:a0b2_base_flow fields}, we illustrate the sub-critical transition from laminar to turbulent state for the uncontrolled and controlled cases with $E_0=1 \times 10 ^{-4}$. 
The energy response of the uncontrolled case~(black solid line) and the SOF-LQR-s controlled case~(red dotted line) exhibit a similar trend.
Transient energy growth reaches a peak value around $t\bar{u}_c/h=60$, then decays until $E/E_0$ reaches approximately $40\%$~(uncontrolled) and $30\%$~(SOF-LQR-s) of its maximum value. During this process~($0 \leq t\bar{u}_c/h\lessapprox 130$), the SOF-LQR-s continuously shows a lower magnitude of $E/E_0$ compared to the uncontrolled flow. 
Additionally, the SOF-LQR-s controller delays the secondary increase in $E/E_0$ from $t\bar{u}_c/h \approx 300$ in the uncontrolled case to $t\bar{u}_c/h \approx 400$.
This secondary increase in perturbation energy in both cases corresponds to a laminar-to-turbulent transition.
Thus, it is evident that the SOF-LQR-s controller delays transition relative to the uncontrolled flow.
In contrast, the LQR controlled flow~(blue dashed line) reaches a comparably lower maximum TEG value earlier in time, but the system is quickly destabilized, and the flow transitions to turbulence.
From the energy responses alone, we can expect for the SOF-LQR-s to delay the laminar-to-turbulent transition, and for the LQR control to promote instability and expedite transition relative to the uncontrolled flow.

To better understand the laminar-to-turbulent transition mechanism, we examine the features of the flow field and divide the entire transition process for each case into several characteristic stages as shown in figure~\ref{fig:a0b2_base_flow fields}~(right). Representative stages in the flow evolution are marked as stage I, II and III. For each stage, the flow snapshot is taken at the same time step for the uncontrolled and the SOF-LQR-s cases. For the LQR case, due to a much earlier transition, the flow field slices are reported for the same stages, but at different time steps.
In all three cases, the initial flow perturbation exhibits streamwise coherent structures. For both the uncontrolled and SOF-LQR-s cases, the flow field remains uniform in the streamwise direction in stage~\rom{1}, and then gradually undergoes spatial distortion after the peak in the kinetic energy density, entering into stage~\rom{2}. Shortly after, rotations start to appear around the streamwise vortical structures with comparable scale to the streamwise domain length. Meanwhile, the kinetic energy density of the flow increases again, initiating stage~\rom{3}, in which the large structures break down and smaller structures can be observed in the flow. After a short duration, the flow transitions into a turbulent state with a rapid increase in $E/E_0$. 
With a similar transition mechanism as the uncontrolled case, the SOF-LQR-s controlled case displays all the three stages of laminar-to-turbulent transition, but their emergence is delayed relative to the uncontrolled case. When the break-down of large structures is observed in the uncontrolled flow, the streamwise vortical structures remain in the SOF-LQR-s as shown at stage~\rom{3}. Hence, the SOF-LQR-s controlled flow results in a delay of the transition. 
The LQR case encounters flow transition with different stages compared to the uncontrolled and the SOF-LQR-s cases. In stage~\rom{1}, the streamwise vortical structures are generated by both the optimal disturbance and the actuation.  A short duration after stage~\rom{1}, secondary instabilities grow in proximity to the walls. The instabilities continuously grow, forming small structures as shown in stage~\rom{2} and later. As time progresses, the small structures become irregular and drive the flow to transition to turbulence, as shown in stage~\rom{3}. A second increase in $E/E_0$ is also observed around $t\bar{u}_c/h=70$.

\begin{figure}
    \centering
    \includegraphics[width=0.6\textwidth]{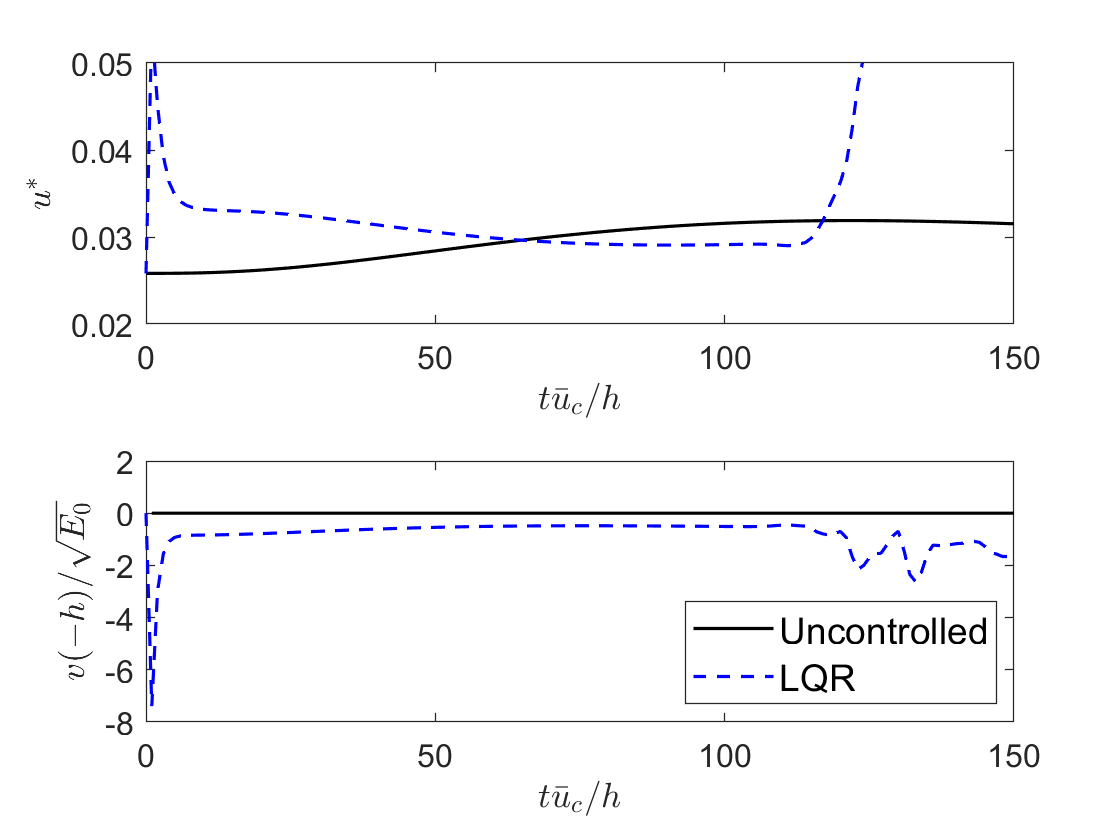}
    \caption{DNS of the uncontrolled flow and the full-information LQR controlled flow with $(\alpha,\beta)=(0,2)$, $Re=3000$, and $E_0=5\times 10 ^{-5}$. (Top) Friction velocity and (Bottom) normalized wall-normal velocity $v(-h)/\sqrt{E_0}$ at the lower channel wall with $x=0$, $z=0$.}
    \label{fig:a0b2_lqr_transition}
\end{figure}

\begin{figure}
    \centering
    \includegraphics[width=0.6\textwidth]{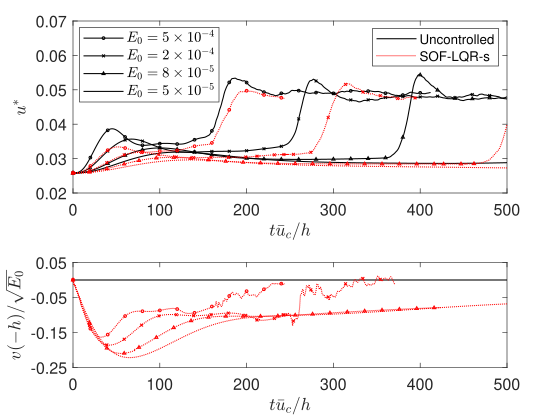}
    \caption{DNS of the uncontrolled flow and the SOF-LQR-s controlled flow with $(\alpha,\beta)=(0,2)$, $Re=3000$, and various perturbation amplitudes $E_0$. (Top) Friction velocity and (Bottom) normalized wall-normal velocity $v(-h)/\sqrt{E_0}$ at the lower channel wall $x=0$, $z=0$.}
    \label{fig:a0b2_sof_transition}
\end{figure}

To examine the transition phenomenon further, we note that a sudden increase in friction velocity $u^*$ has been used as evidence for transition in channel flow, since turbulent boundary layers tend to have larger shear-stress at the walls than laminar ones. The friction velocity is defined as, 
\begin{equation}
    u^*=\sqrt{\frac{{\tau}_w}{\rho}},
\end{equation}
where ${\tau}_w$ is the wall shear-stress, and $\rho$ is the density of the fluid.
The transition events in the uncontrolled and controlled flows are illustrated by a sharp increase of the friction velocity $u^*$ as shown in figures~\ref{fig:a0b2_lqr_transition} and~\ref{fig:a0b2_sof_transition}.
Interestingly, although the LQR controller reduces TEG to a great extent, it causes a decrease in the transition threshold. As shown in figure~\ref{fig:a0b2_lqr_transition}, the uncontrolled flow stays in the laminar regime with $E_0=5\times 10 ^{-5}$ when the LQR controlled flow has already transitioned to turbulence. The LQR controller gives rise to an initial peak of the friction velocity and a sudden drop in the normalized lower wall velocity. This is due to the relatively high control input introduced by the LQR controller, which also relates to the earlier onset of transition in this case, as will be discussed in detail momentarily. 
In contrast, with less reduction in the TEG, the SOF-LQR-s controller delays the laminar-to-turbulent transition compared to the uncontrolled flow, as shown in figure~\ref{fig:a0b2_sof_transition}. The delay is more significant under lower kinetic energy density amplitudes.
Moreover, although transition still arises in the controlled flow, the friction velocity is significantly reduced compared to the uncontrolled flow, which ultimately yields a drag reduction in the controlled flow. 
The control input generated by the SOF-LQR-s has a similar trend as the normalized lower wall velocity history. Compared to the LQR case, the SOF-LQR-s control input is much smaller and avoids rapid changes, which benefits the delaying of transition.

\begin{figure}
    \centering
    \includegraphics[width=0.75\textwidth]{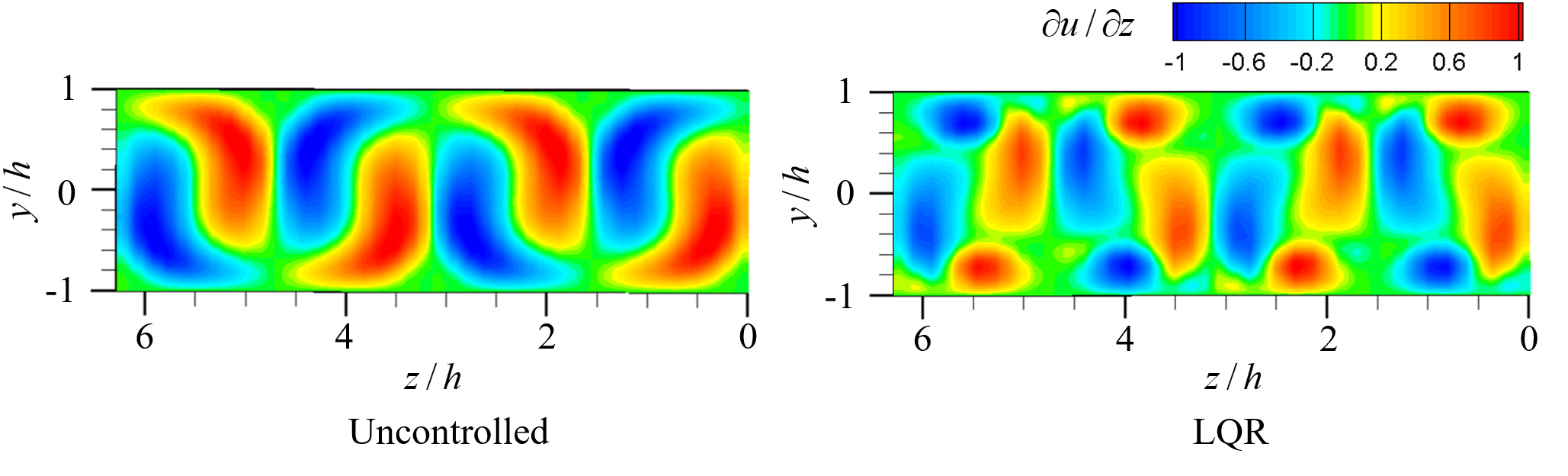}
   \caption{Comparison between uncontrolled flow~(left) and LQR controlled flow~(right) with ($\alpha,\beta$) = (0,2), $Re=3000$. Modification of instantaneous streamwise velocity gradient in spanwise direction $\partial{u}/\partial{z}$ at $t\bar{u}_c/h=49$, $x/h$=0.m}
    \label{fig:a0b2_sof_dudz}
\end{figure}

\begin{figure}
\centering
\subfloat[Slice at $t\bar{u}_c/h=49$, $x/h=0$]{\label{fig:a0b2_sof_shear49}
\includegraphics[width=0.75\textwidth]{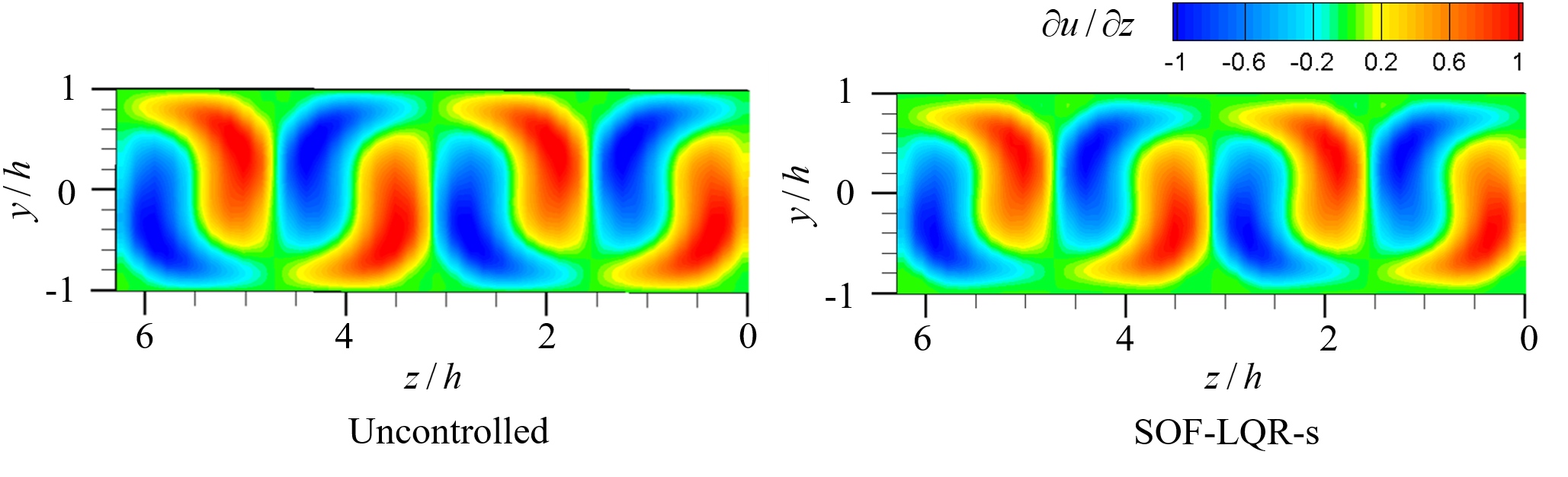}
} \\
\vspace{1pt}
\subfloat[Slice at $t\bar{u}_c/h=120$, $x/h=0$]{\label{fig:a0b2_sof_shear120}
\includegraphics[width=0.75\textwidth]{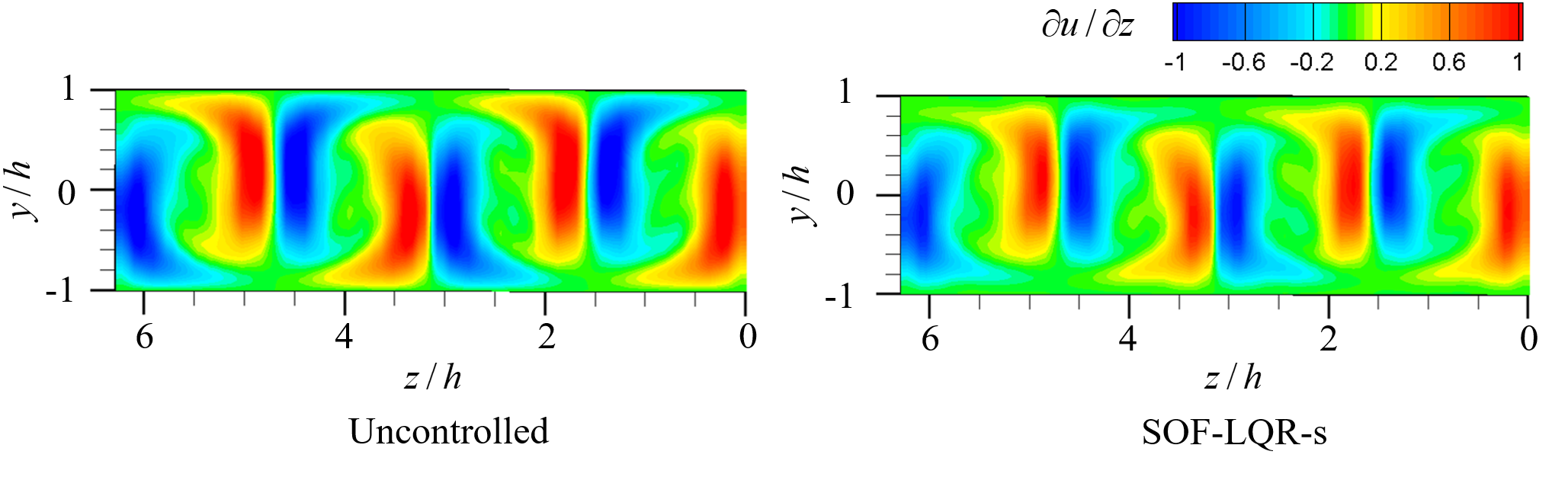}
} 
\caption{Comparison between uncontrolled flow~(left) and SOF-LQR-s controlled flow~(right) with ($\alpha,\beta$) = (0,2), $Re=3000$. Modification of instantaneous streamwise velocity gradient in spanwise direction $\partial{u}/\partial{z}$ at $t\bar{u}_c/h=49$, $x/h=0$~(a) and $t\bar{u}_c/h=120$, $x/h=0$~(b).} \label{fig:a0b2_sof_shear}
\end{figure}

From the observations above, it appears that the onset of transition cannot be reliably determined from an examination of the TEG alone. 
Consider that  the LQR reduces TEG to less than a quarter of the TEG for the uncontrolled flow. Yet, the LQR controller drives the flow to transition even earlier than the baseline uncontrolled case.
This phenomenon has been closely examined in our previous study~\cite{sun2019}. As shown in figure~\ref{fig:a0b2_sof_dudz}, the actuation introduced by the LQR controller forms small-scale streamwise vortices near the walls and suppresses transient energy growth by weakening the growth of the large streamwise vortical structures in the uncontrolled flow. However, the generation of extra high-shear regions introduces secondary instabilities and drives the flow to transition. 
Comparably, the SOF-LQR-s reduces TEG to about half of the baseline case. The reduction is less than what is achieved by the LQR control; however, this is sufficient to successfully delay the transition. By considering slices of the flow field (see figure~\ref{fig:a0b2_sof_shear}), it becomes apparent that the SOF-LQR-s does not change the shape of the coherent structures, but does shrink their size and reduce the shear-stress amplitude.
Notably, it decreases the magnitude of the shear-stress in the vicinity of the walls. This is more clearly observed in the later stages (e.g., $t\bar{u}_c/h=120$ in figure~\ref{fig:a0b2_sof_shear120}). Also, while limiting the size of coherent structure, the control input is not so large as generate additional small-scale vortices near the walls, in contrast to the LQR control case. Thus, we observe both TEG reduction and transition delay in the SOF-LQR-s controlled flow.

Note that for controller design in this study, we have opted to design the LQR and SOF-LQR-s without penalization on the control inputs.
That is, the weighting matrix $R$ in~\eqnref{eqn:objective_function} and~\eqnref{eqn:sof_obj} is chosen to be small~($R=1\times 10^{-6}I$) in order to achieve the best regulation possible without regard for the control effort. However, considering that the enhanced transition mechanism in LQR control is due to the heavy-handed actuation, we can modify the design objective to penalize the control input for achieving improved nonlinear transition performance. 
Indeed, it is possible to design the LQR controller to regulate the flow while preventing the appearance of additional large-shear areas near the walls, which are otherwise introduced by the control input. This would be done by sacrificing the large TEG reduction for a more moderate reduction.
As such, the results above do not suggest that the full-state feedback LQR will always fail for transition control.  Rather, we include these results to emphasize that the ``right'' objective for transition control must strike a balance between reducing TEG and avoiding secondary instabilities caused by the control input.

\subsection{Streamwise disturbances}
\subsubsection{Linear analysis}
\begin{figure}[h!]
\centering
\includegraphics[width=0.475\textwidth]{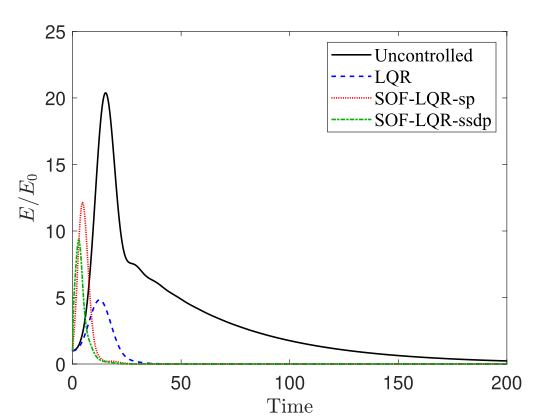}
\caption{Linear worst-case response to streamwise optimal perturbations with $(\alpha,\beta)=(1,0)$ and $Re=3000$.}
\label{fig:a1b0_TEG}
\end{figure}

For streamwise disturbances $(\alpha,\beta)=(1,0)$, we investigate SOF-LQR control using sensor configurations ``sp''~(i.e. $\tilde{\tau}_{{x}},\tilde{\tau}_{{z}},\tilde{p}$) and ``ssdp"~(i.e. $\tilde{\tau}_{{x}},\tilde{\tau}_{{z}},  \frac {\partial \tilde{\tau}_{{x}}}{\partial y},\frac {\partial \tilde{\tau}_{{z}}}{\partial y} ,\tilde{p}$) as outlined in Table~\ref{tab:sensors}.
Although not reported here, SOF-LQR-s controllers---using only $\tilde{\tau}_{{x}}$ and $\tilde{\tau}_{{z}}$ sensors for feedback---were found to increase the maximum TEG arising from streamwise optimal disturbances.
The inclusion of pressure information along with the shear-stress sensors (i.e.,~configuration ``sp'') was found to be important for TEG suppression. Moreover, access to additional information pertaining to spatial derivatives of shear-stress with respect to the wall-normal direction (i.e.,~configuration ``ssdp'') was found to improve TEG reduction even further.
In the linear simulations of the worst-case response for $(\alpha,\beta)=(1,0)$ and $Re=3000$ (see Figure~\ref{fig:a1b0_TEG}), the full-information LQR controller~(dashed blue line) reduces the maximum TEG~($G$) by $76\%$ relative to the uncontrolled flow~(black solid line).
The SOF-LQR controllers also reduce the maximum TEG relative to the uncontrolled flow, but not to the same extent:
SOF-LQR-sp~(red dotted line) reduces TEG by $41\%$  
and SOF-LQR-ssdp~(green dash-dotted line) by $54\%$.
Despite the superior linear performance of SOF-LQR-ssdp, we will discuss both sets of sensor configurations in the remaining analysis in order to shed light on the role of various sensed quantities in transition control.

\begin{figure}
\subfloat[Controllers designed for $(\alpha,\beta)=(1,0)$]{\label{fig:robustness_a1b0}
\includegraphics[width=0.4\textwidth]{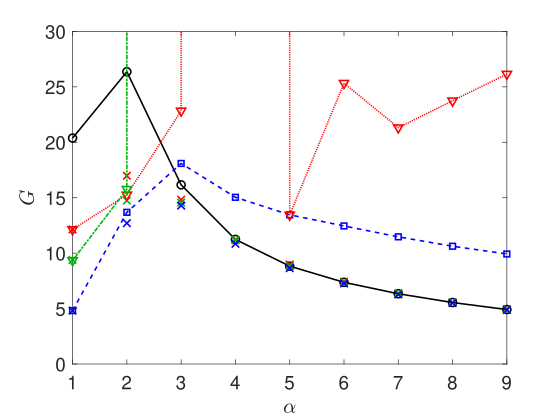}
} 
\hspace{1pt}
\subfloat[Controllers designed for $(\alpha,\beta)=(3,0)$]{\label{fig:robustness_a3b0}
\includegraphics[width=0.4\textwidth]{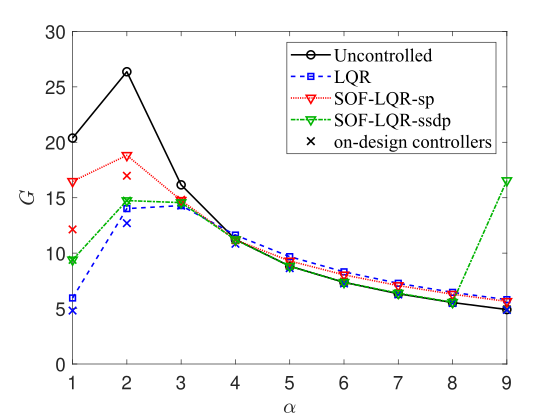}
}
\caption{Robustness of LQR and SOF-LQR controllers to off-design streamwise perturbations at $Re=3000$.} \label{fig:streamwise_robustness}
\end{figure}

Next, we consider the robust performance of SOF-LQR controllers at off-design conditions.
As with the case of robust control of spanwise disturbances, we will see that the choice of on-design conditions can have a remarkable impact on robust control performance.
We begin by considering controllers designed for the on-design condition of $(\alpha,\beta)=(1,0)$ and $Re=3000$, then apply these controllers to streamwise optimal disturbances at off-design conditions with $\alpha=[1,9]$, $\beta=0$, and $Re=3000$ (see Figure~\ref{fig:robustness_a1b0}).
Although the SOF-LQR-ssdp controller reduces TEG to a higher-degree at this on-design condition, the same controller proves to be quite fragile to off-design streamwise perturbations with $\alpha>2$.
In fact, worst-case analysis at these off-design conditions shows that SOF-LQR-ssdp designed for $(\alpha,\beta)=(1,0)$ will linearly destabilize the flow when subjected to these higher wavenumber streamwise perturbations.
In contrast, the SOF-LQR-sp controller designed at the same on-design condition at least guarantees closed-loop stability for off-design perturbations, except when subjected to optimal streamwise perturbations with $\alpha=4$.
Nonetheless, the robust performance characteristics of SOF-LQR-sp with regards to TEG at other off-design streamwise wavenumbers tends to be larger than that of the uncontrolled flow.
This is also true for the full-information LQR controller when subjected to off-design streamwise disturbances, but to a lesser degree than either SOF-LQR controller.

Based on the analysis above, it seems that SOF-LQR strategies lack robustness to off-design conditions; however, this is not necessarily the case.
Consider controllers designed at the on-design condition of $(\alpha,\beta)=(3,0)$ and $Re=3000$ (see Figure~\ref{fig:robustness_a3b0}).
In this case, all of the TEG control strategies exhibit robust performance when subjected to off-design streamwise optimal disturbances---except for SOF-LQR-ssdp at $\alpha=9$.
When subjected to off-design streamwise disturbances with $\alpha> 3$, the full-information LQR and SOF-LQR-sp controllers result in a (moderately) larger TEG than the uncontrolled flow.
Further, when subjected to off-design streamwise disturbances with $\alpha<3$, the full-information LQR and SOF-LQR-sp controllers have moderately larger TEG than what would be achieved by designing the respective controllers for these specific disturbances~(marked as crosses).
In contrast, the SOF-LQR-ssdp controller designed at $(\alpha,\beta)=(3,0)$ matches its worst-case on-design TEG performance at off-design conditions (except when $\alpha=9$).
For mitigation of TEG due to  streamwise disturbances, we observe a similar phenomenon as in the spanwise disturbance mitigation scenario: Controllers designed for higher streamwise wavenumber disturbances exhibit better robustness properties. These controllers yield linear stable closed-loop dynamics at off-design streamwise wavenumbers lower than the on-design one, and to a larger range of off-design streamwise wavenumbers above the on-design one (i.e.,~robust stability). 
 However, off-design TEG performance tends to be degraded when subjected to higher wavenumber disturbances (i.e.,~robust performance), but this can be avoided with the choice of on-design condition.
This analysis suggests that on-design conditions for controller synthesis are important to robust TEG control, both for full-information and sensor-based output feedback strategies.

\begin{figure}
\includegraphics[width=0.475\textwidth]{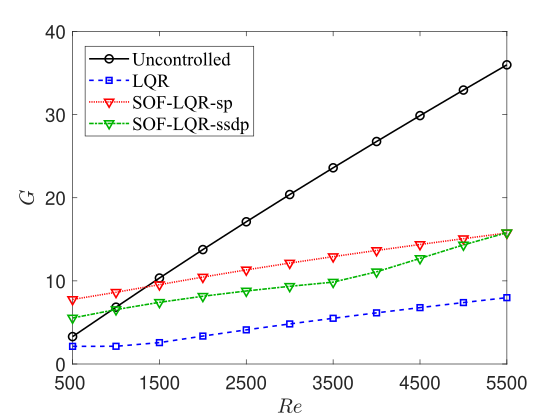}
\caption{Robustness of LQR and SOF-LQR controllers to off-design Reynolds numbers. Both controllers are designed for $Re=3000$, but robustly reduce TEG from streamwise optimal perturbations over a range of off-design Reynolds numbers.} \label{fig:a1b0_MTEG_robustness}
\end{figure}

We additionally assess streamwise controller robustness to Reynolds number variations.
To do so, we again consider $(\alpha,\beta)=(1,0)$ and $Re=3000$ as the on-design condition, then apply the resulting  controllers at ``off-design'' sub-critical Reynolds number conditions in the range, $Re=[500,5500]$ (see Figure~\ref{fig:a1b0_MTEG_robustness}).
 For off-design $Re\geq 1500$, the full-information LQR controller and both SOF-LQR controllers reduce TEG relative to the uncontrolled flow. 
At lower off-design $Re$, the SOF-LQR controllers give rise to moderately higher TEG than without control. This occurs at $Re<1500$ for SOF-LQR-sp and at $Re<1000$ for SOF-LQR-ssdp.
We conclude that the additional information about spatial derivatives of shear-stress with respect to the wall-normal direction improves TEG reduction and---when designed properly---robustness properties of the SOF-LQR controller.
Thus, it can be expected that SOF-LQR-ssdp will outperform SOF-LQR-sp for transition control in nonlinear direct numerical simulations, which we investigate next. 
\subsubsection{Nonlinear direct numerical simulation}
\begin{figure}
    \centering
    \includegraphics[width=0.6\textwidth]{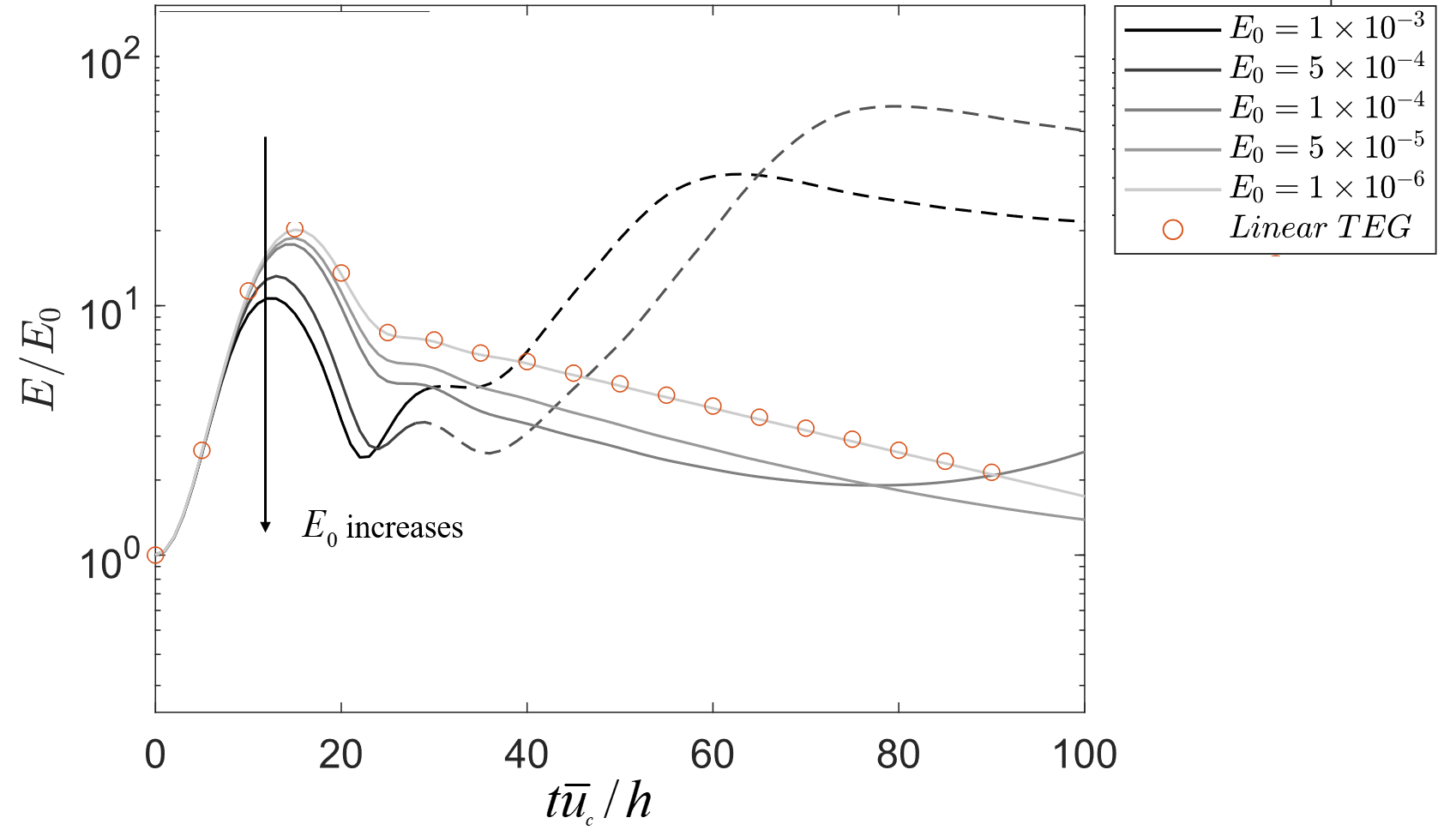}
    \caption{($\alpha,\beta$) = (1,0) transient energy growth of optimal disturbance in uncontrolled flow with various amplitudes of $E_0$.}
    \label{fig:a1b0_base_E}
\end{figure}

\begin{figure}
    \includegraphics[width=0.95\textwidth]{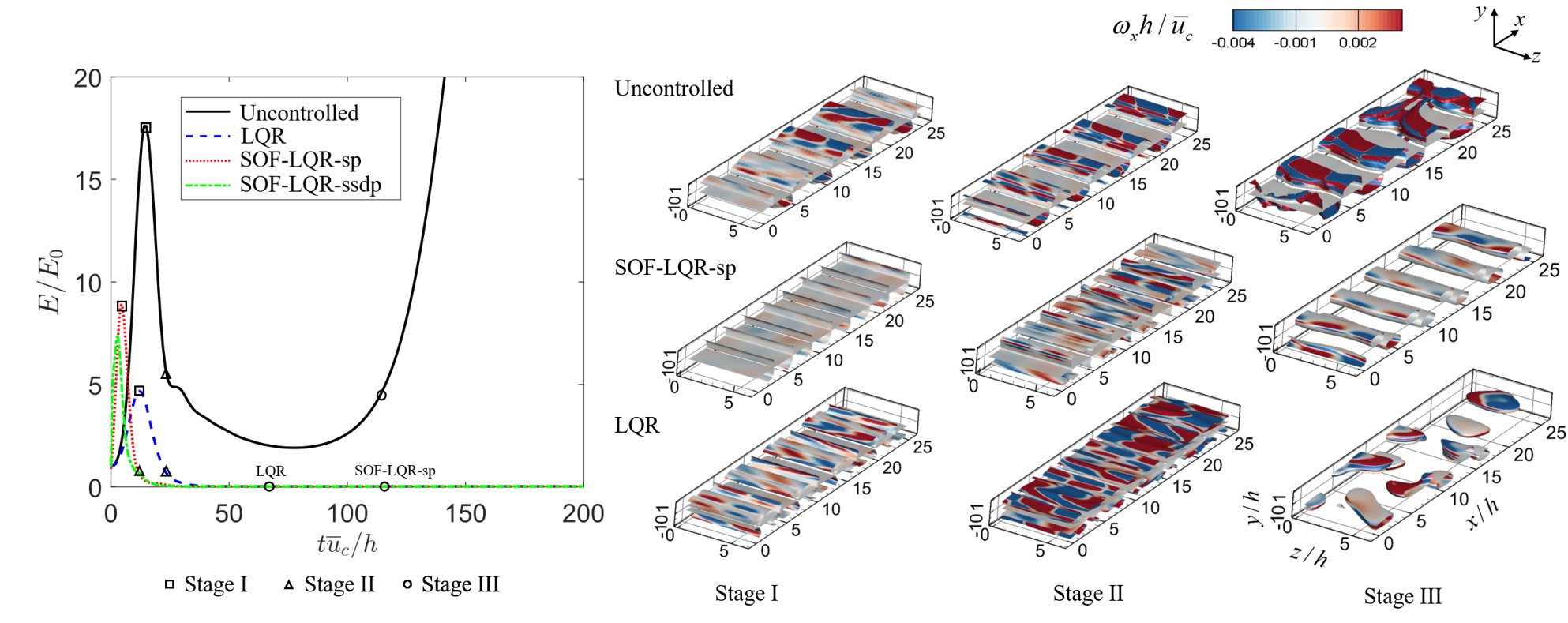}
    \caption{Transient energy growth energy and friction velocity on walls of uncontrolled flow ($E_0 =1\times 10^{-4}$). Inserts are corresponding iso-surface of $\mathcal{Q}$-criterion \citep{Hunt:88}~($\mathcal{Q}(h/\bar{u}_c)^2=1\times 10^{-4}$) colored by streamwise vorticity $\omega_x h/\bar{u}_c$.}
    \label{fig:a1b0_flow fields}
\end{figure}

Direct numerical simulations are performed under the linear optimal streamwise disturbance with $(\alpha,\beta)=(1,0)$ and $Re=3000$.
Results in figure~\ref{fig:a1b0_base_E} show that when the kinetic energy density amplitude is $E_0=1\times 10 ^{-6}$, the nonlinear flow response overlaps with the linear result~(see red circles).
The uncontrolled flow transitions to turbulence at an amplitude threshold of $E_0 = 1 \times 10^{-4}$, as shown in figure~\ref{fig:a1b0_base_E}.
In figure~\ref{fig:a1b0_flow fields}, we report the results for the uncontrolled flow and the controlled flow in response to linear optimal disturbances with $E_0=1\times 10^{-4}$.
The flow fields at different time stages illustrate the transition mechanism in the uncontrolled flow and the transition suppression mechanisms in the controlled flows. In figure~\ref{fig:a1b0_flow fields}, representative stages are marked on the energy plot on the left with the corresponding flow fields illustrated on the right. Stages I and II correspond to the time that $E/E_0$ reaches its peak and when $E/E_0$ decays after this peak, respectively. The uncontrolled and LQR flow snapshots for stages I and II correspond to the same convective times. The SOF-LQR-sp stages arise at different times due to an earlier appearance of the $E/E_0$ peak. Stage III corresponds to transition to turbulence~(uncontrolled), or to the return to the laminar regime~(controlled).

As shown in figure~\ref{fig:a1b0_flow fields} on the left, the uncontrolled flow reaches a maximum transient energy growth of $E/E_0 \approx 17$ at $t{\bar{u}}_c/h \approx 15$.
For the controlled cases, the full-information LQR controlled flow reaches maximum transient energy growth around a similar time, but with a reduced amplitude of $E/E_0 \approx 5$.
The representative flow fields show that the initial optimal streamwise disturbance grows into large-scale spanwise coherent structures. 
In the uncontrolled flow, streamwise vorticity increases near the channel walls, leading to the formation of $\Lambda$-shaped structures (see stage \rom{3}) and a transition to turbulence shortly after.
In the full-information LQR controlled flow, the spanwise coherent structures remain separate from stage \rom{1} to \rom{2}.
Although high-shear regions can be observed in stage \rom{2}, these large structures quickly decay and break-down into smaller isolated structures in stage \rom{3}.
These small structures eventually decay completely, and the flow remains laminar.
In the SOF-LQR controlled flows---using either sensor configuration---the maximum transient energy growth occurs at an earlier time than under full-information LQR control.
For SOF-LQR-sp, the maximum transient energy growth reaches $E/E_0 \approx 9$ at $t{\bar{u}}_c/h \approx 5$, and for SOF-LQR-ssdp it reaches $E/E_0 \approx 7$ at $t{\bar{u}}_c/h \approx 3$.
The transition suppression scenarios are similar between the two SOF-LQR controllers, and so we use SOF-LQR-sp to demonstrate the flow field evolution.
For the SOF-LQR-sp controlled flow, the spanwise coherent structures that evolve out of the optimal streamwise disturbance persist until a later stage than for the full-information LQR controlled case.
From stage \rom{1} to stage \rom{3}, the spanwise coherent structures decay while remaining intact, not breaking apart as in the case of full-information LQR control. 
The streamwise vorticity magnitude also tends to be smaller compared to either the uncontrolled flow or the full-information LQR controlled flow. The large spanwise coherent structures in SOF-LQR-sp persist for a longer time compared to the full-information LQR case, but do ultimately fully decay.
However, the SOF-LQR control is able to prevent the formation of $\Lambda$-structures that would cause the spanwise coherent structures to merge, and thus successfully suppresses transition to turbulence.  

\begin{figure}
\centering
\subfloat[LQR]{\label{fig:a0b2_lqr_frictionv}
\includegraphics[width=0.48\textwidth]{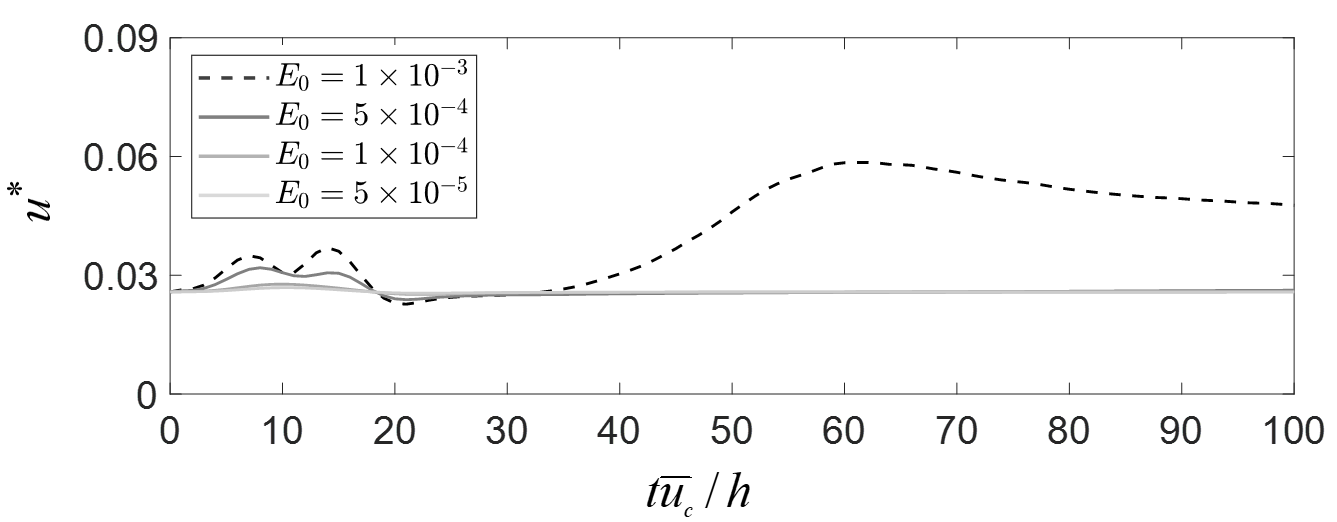}
\includegraphics[width=0.48\textwidth]{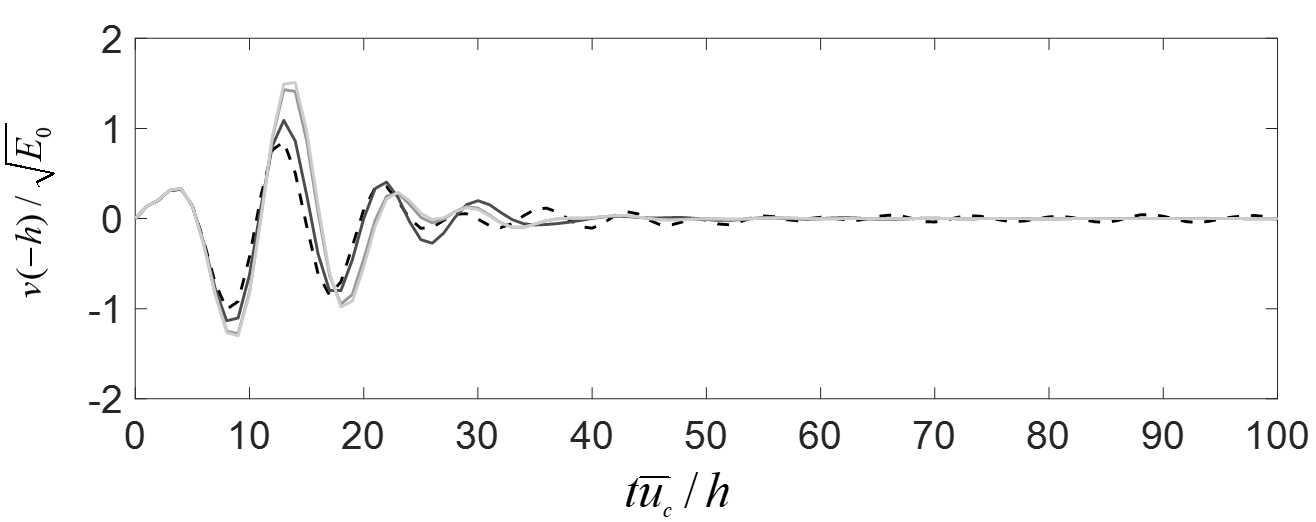}
} \\
\vspace{1pt}
\subfloat[SOF-LQR-sp]{\label{fig:a0b2_sof1_frictionv}
\includegraphics[width=0.48\textwidth]{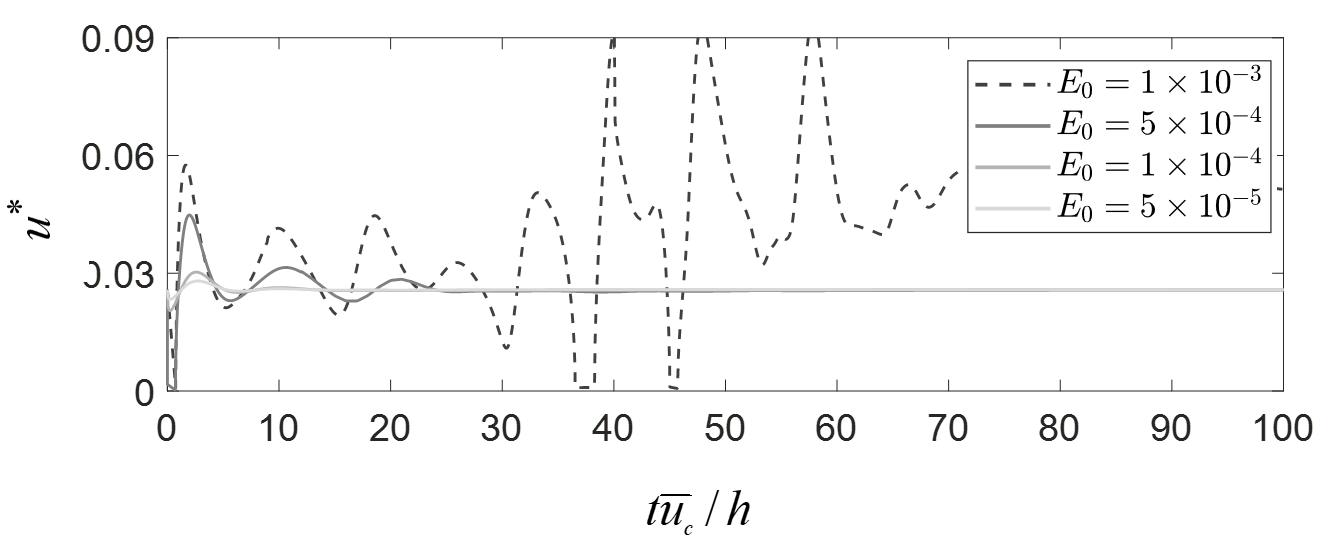}
\includegraphics[width=0.48\textwidth]{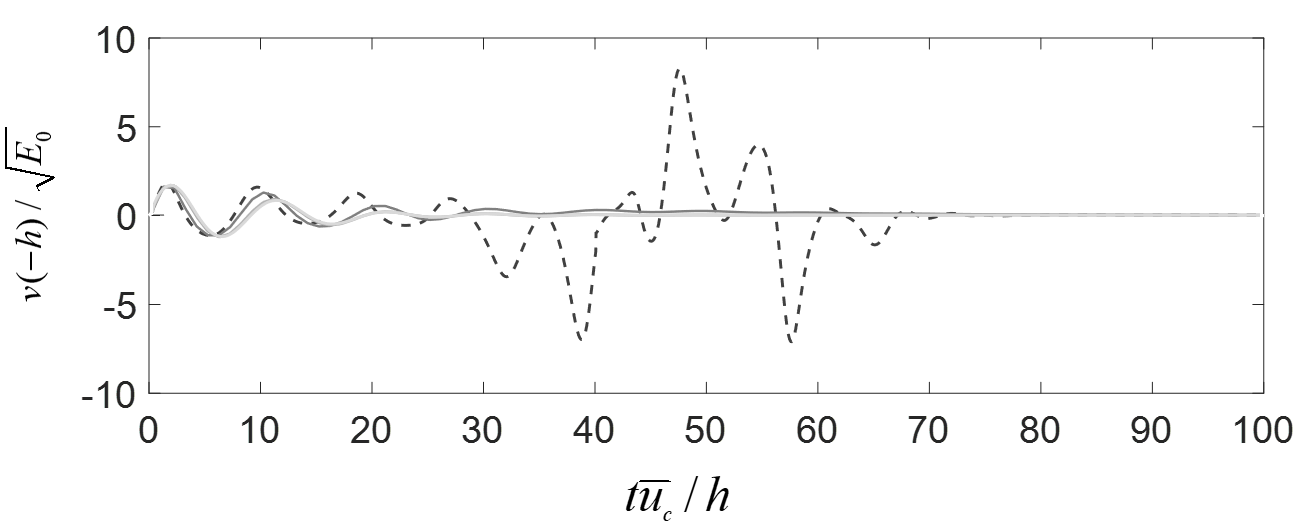}
}
\vspace{1pt}
\subfloat[SOF-LQR-ssdp]{\label{fig:a0b2_sof2_frictionv}
\includegraphics[width=0.48\textwidth]{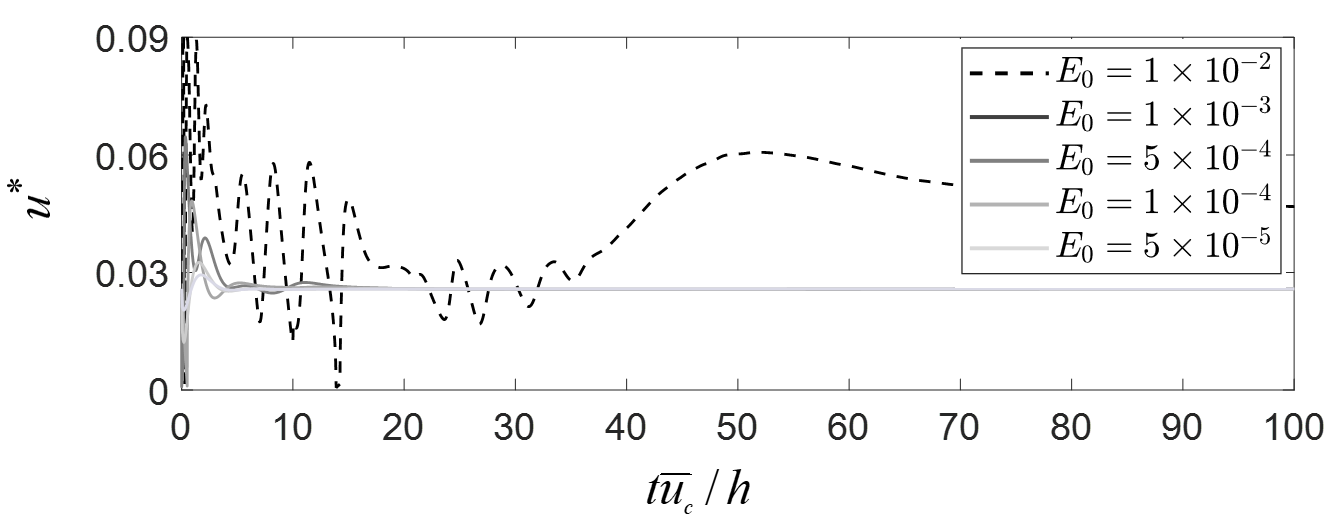}
\includegraphics[width=0.48\textwidth]{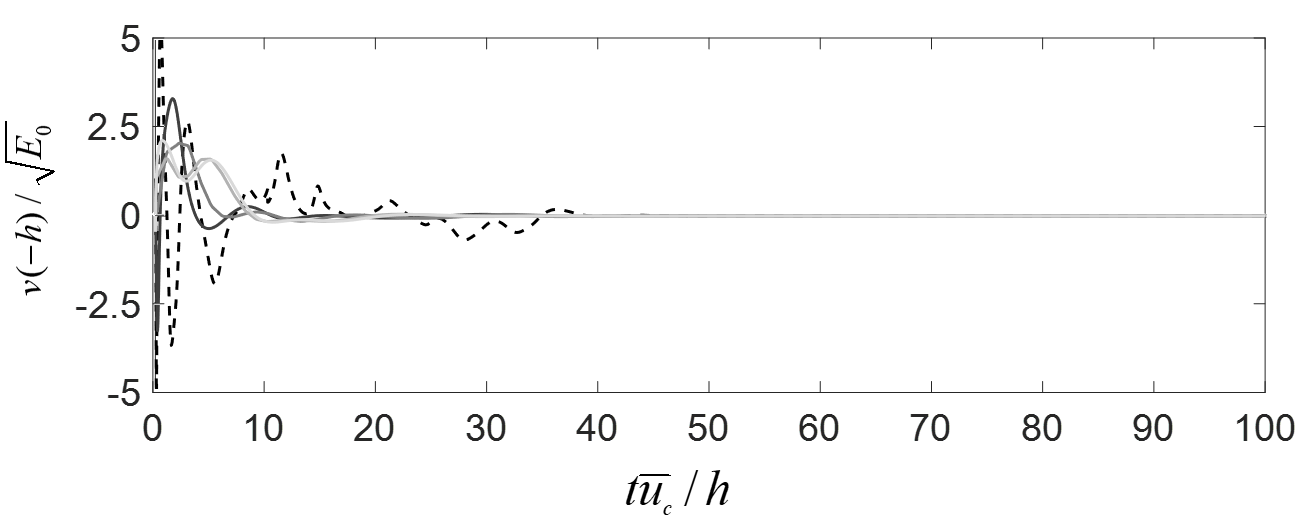}
}
\caption{DNS of the uncontrolled flow and the controlled flow with $(\alpha,\beta)=(1,0)$, $Re=3000$, and various perturbation amplitudes $E_0$. (Left) Friction velocity and (Right) normalized wall-normal velocity $v(-h)/\sqrt{E_0}$ at the lower channel wall.} \label{fig:a1b0_friction_velocity}
\end{figure}

The time-histories of friction velocity and normalized wall-normal velocity at the lower wall for the uncontrolled and controlled flows are shown in figure~\ref{fig:a1b0_friction_velocity}. 
By analyzing different perturbation amplitude levels, we see that the full-information LQR controller increases the transition threshold from $E_0=1\times 10^{-4}$ of the uncontrolled flow to $E_0=1\times 10^{-3}$, as shown in figure~\ref{fig:a1b0_friction_velocity} and as reported  in our previous work~\cite{sun2019}.
The SOF-LQR-sp controller also increases the transition threshold to $E_0=1\times 10^{-3}$, whereas the SOF-LQR-ssdp increases the transition threshold by another order of magnitude to $E_0=1\times 10^{-2}$.
Although all controllers increase the transition threshold, we again point out that TEG reduction is not a reliable predictor of transition control performance. 
We observe that the SOF-LQR-ssdp controller reduces TEG to a greater degree than the SOF-LQR-sp controller, and that it also increases the transition threshold by an additional order of magnitude.
However, based on TEG reduction performance alone, one would expect the full-information LQR controller to exhibit the best transition control performance. Yet, this is not the case that we observe in nonlinear simulations with the given controller designs.

\begin{figure}
\centering
\subfloat[Slices at convective time $t{\bar u}_c/h=10$ for uncontrolled and controlled flow]{\label{fig:a1b0_contours_1}
\includegraphics[width=0.7\textwidth]{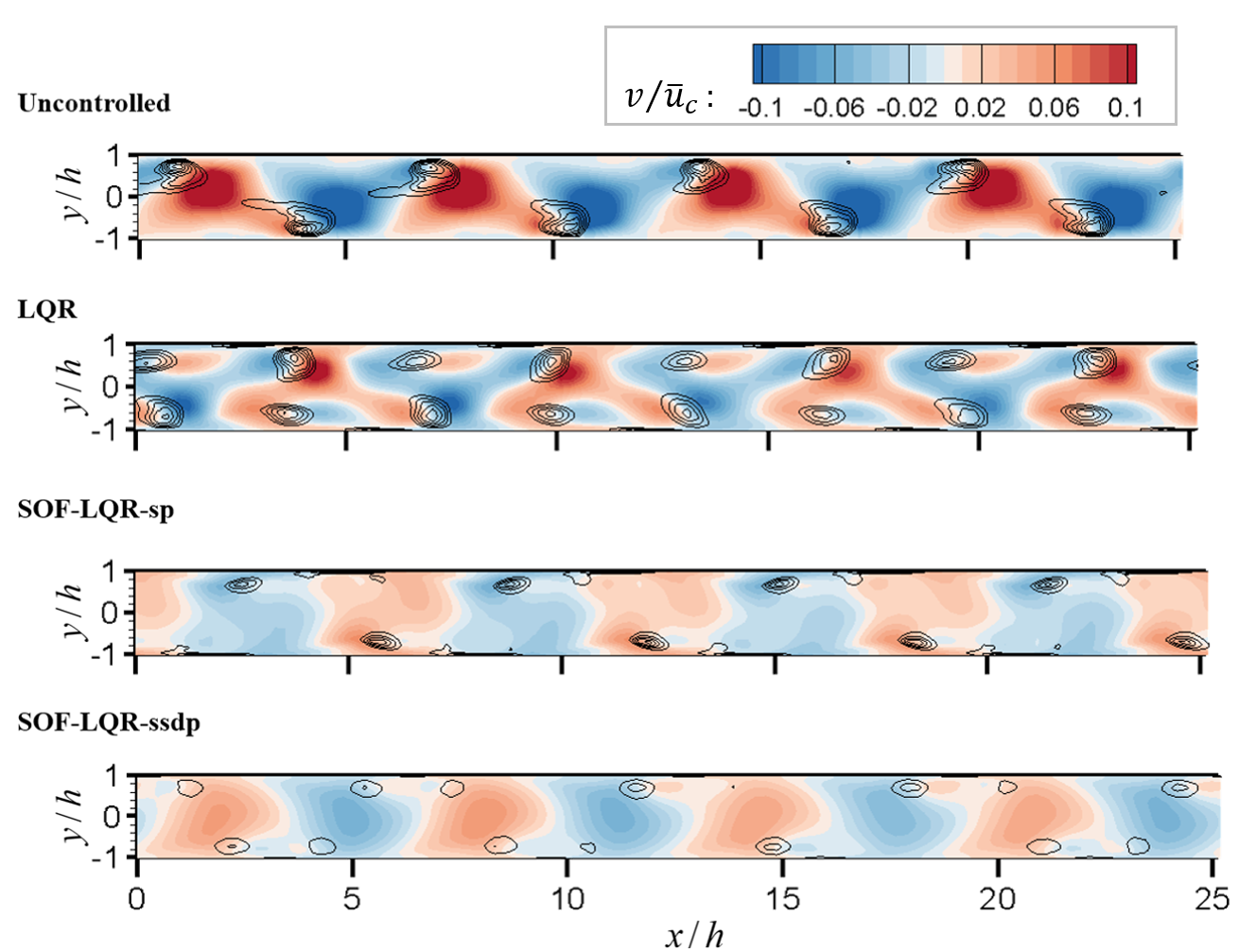}
} \\
\subfloat[Slice at convective time $t{\bar u}_c/h=3$ for SOF-LQR controllers]{\label{fig:a1b0_contours_2}
\includegraphics[width=0.7\textwidth]{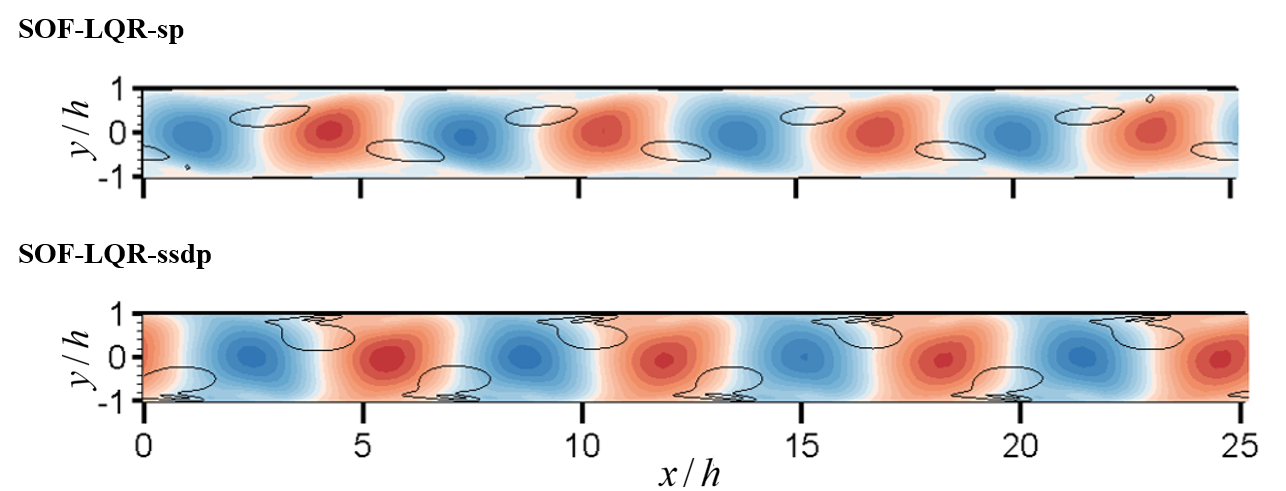}
}
\caption{Flow fields of $(\alpha,\beta)=(1,0)$ with $E_0=1 \times 10^{-3}$ for uncontrolled and controlled flow. Contours are wall-normal velocity, and black contour lines denotes $\mathcal{Q}$-criterion in a range of $[0.05, ~1]$.}
\label{fig:a1b0_contours}
\end{figure}

All the controllers modify the uncontrolled flow features through wall normal actuation.
The actuated wall-normal velocity histories are shown in figure~\ref{fig:a1b0_friction_velocity}~(on the right). 
The controllers generate large actuation to suppress TEG over a relatively short transient time-horizon of $0<t\bar{u}_c/h<30$.
For the small amplitude disturbances, when the controller is able to reduce the TEG as well as suppress transition, the actuation gradually settles to zero.
When facing the large amplitude disturbances, the actuation fails to prevent the flow from transitioning to turbulence, and so the actuation will oscillate for a longer time associated with the disturbance phase speed and wavenumber~\citep{sun2019}.
In figure~\ref{fig:a1b0_contours}, we present $x$-$y$ slices of the uncontrolled and the controlled flow fields along with coherent structures visualized using the $\mathcal{Q}$-criterion. 
The first group of slices is extracted at convective time $t{\bar u}_c/h=10$, as shown in figure~\ref{fig:a1b0_contours_1}.
At this time, both the uncontrolled flow and LQR controlled flow reach a maximum $E/E_0$. 
Since the SOF-LQR controllers reach the $E/E_0$ peak at an earlier time, we additionally take a slice from each of the SOF-LQR controllers at $t{\bar u}_c/h=3$, as shown in figure~\ref{fig:a1b0_contours_2}.
Compared to the uncontrolled flow, all of the controllers decrease the wall-normal velocity and limit the size of the coherent structures when the normalized perturbation kinetic energy density reaches its peak value. 
This reduction also decreases the likelihood for break-down of streamwise vortical structures caused by the large spanwise coherent structures discussed in the context of figure~\ref{fig:a1b0_flow fields}.
As a result, the controllers are able to successfully delay or prevent laminar-to-turbulent transition. 

As noted earlier, even though the full-information LQR controller yields superior performance for TEG reduction, the highest transition threshold is actually provided by the SOF-LQR-ssdp controller.
A comparison of the flow field slices reveals that both SOF-LQR controllers reduce velocity fluctuations and growth of large coherent structures to a greater extent than the full-information LQR controller.
Comparing the two SOF-LQR controllers, it is the one with more sensors that reduces TEG to a greater extent and exhibits superior performance in preventing transition in the nonlinear simulations. 
From this analysis, we see that the reduction of the linear TEG is useful in guiding controller design, but is not the only metric that should be used when considering laminar-to-turbulent transition control.

\section{Conclusion} \label{sec:conclusions}
In this paper, we investigated SOF-LQR controllers for reducing transient energy growth~(TEG) of flow perturbations in a channel flow with wall-based sensing and actuation. 
Control was achieved using wall-normal blowing and suction actuation. Several wall-based sensor combinations were investigated, including shear-stress, wall-normal derivative of shear-stress, and pressure.
SOF-LQR controllers were designed using an Anderson-Moore algorithm.
These computations were accelerated by leveraging Armijo-type
adaptations to reduce the iteration count and expedite calculations.

SOF-LQR controllers were found to reduce the worst-case TEG relative to the uncontrolled flow in linear simulations.
Spanwise perturbations are associated with the largest levels of TEG for the linearized channel flow, and TEG reduction can be achieved using shear-stress measurements alone.
The designed SOF-LQR controllers were found to reduce TEG under streamwise perturbations with shear-stress and pressure measurements. Including measurements of the wall-normal derivative of shear-stress was found to further reduce TEG in this case. 
 All SOF-LQR controllers investigated here exhibited robustness to Reynolds number uncertainties.
 Robustness to wavenumber uncertainties was also evaluated.  We found that designing SOF-LQR controllers at a proper on-design condition enabled robust performance to these modeling uncertainties at off-design conditions based on linear worst-case analysis.

In nonlinear direct numerical simulations, SOF-LQR controllers were also found to suppress and delay transition to turbulence, increasing transition thresholds relative to the uncontrolled flow. 
For spanwise disturbances, SOF-LQR control was found to delay transition relative to the uncontrolled flow.
In the case of streamwise disturbances, SOF-LQR control suppressed transition and increased the transition threshold for the uncontrolled flow by up to two orders of magnitude.
The results show that SOF-LQR controllers provide a simple and reliable alternative to other sensor-based output feedback control strategies for TEG reduction and transition control.
The encouraging results presented here motivate further investigation of static output feedback control in more complex flows in the future.
 
 \section{Acknowledgements}
 This material is based upon the work supported by the Air Force Office of Scientific Research under award number FA9550-19-1-0034, monitored by Dr. Gregg Abate.
\appendix
\section{Stabilizing static output feedback controller} \label{sec:app1}
As discussed in section~\ref{Sec:feedbacks}, an initial stabilizing static output feedback~(SOF) gain $F_0$ is necessary for the calculation of the optimal SOF-LQR controller based on the accelerated Anderson-Moore algorithm~(Algorithm~\ref{Algorithm1}). Here, we summarize the iterative linear matrix inequality~(LMI) algorithm from~\citep{Cao1998a} used for determining such a stabilizing SOF controller in this study.

A stabilizing SOF controller gain can be solved using Lyapunov-based control synthesis methods.  
This problem can formulated as an LMI feasibility problem, which can be solved using standard methods.
A stabilizing SOF gain $F_0$ must satisfy the following Lyapunov inequality,
\begin{equation}
   (A+BF_0C)\trans P+P(A+BF_0C)<0. \label{eqn:A1}
\end{equation}
However, this problem is not linear in the design variables $F_0$ and $P$ as written.
Next, recognize that the following two conditions are sufficient conditions for the inequality above to hold:

\textbf{Sufficient condition 1}
\begin{equation}
\begin{aligned}
     &(A+BF_0C)\trans P+P(A+BF_0C) \\ < &(A+BF_0C)\trans P+P(A+BF_0C)+C\trans F_0\trans F_0C\\
     = &A\trans P+PA-PBB\trans P+(B\trans P+F_0C)\trans (B\trans P+F_0C)\\<&0
     \end{aligned}
\end{equation}

\textbf{Sufficient condition 2}
\begin{equation}
\begin{aligned}
&A\trans P+PA-PBB\trans P+(B\trans P+F_0C)\trans (B\trans P+F_0C)\\ < &A\trans P+PA \underbrace{+X\trans BB\trans X-P\trans B B\trans X-X\trans BB\trans P}_{\geq -PBB\trans P} \\+&(B\trans P+F_0C)\trans (B\trans P+F_0C)\\<&0
     \end{aligned}
\end{equation}
where
\begin{equation}
\begin{aligned}
(X-P)\trans BB\trans(X-P) &\geq 0 \\
X\trans BB\trans X-P\trans BB\trans X-X\trans BB\trans P &\geq -PBB\trans P.
\end{aligned}
\end{equation}
Thus, the original problem~\eqnref{eqn:A1} can be recast as the following LMI feasibility problem for a stabilizing SOF gain $F_0$:
\begin{equation}
    \begin{aligned}
   & \left[ \begin{array}{cc}
         A\trans P+PA+X\trans BB\trans X-P\trans BB\trans X-X\trans BB\trans P & (B\trans P+F_0C)\trans \\
        (B\trans P+F_0C) & -I
    \end{array} \right] <0 \\
    & P=P\trans >0.
    \end{aligned}
\end{equation}
Further details regarding this formulation can be found in~\cite{Cao1998a}.

The SOF gain $F_0$ can be determined using off-the-shelf interior point solvers; however, the computational cost of solving this problem scales with $\mathcal{O}(n^6)$, where $n$ is the state dimension.
In this work, we overcome the issue with computational complexity by taking advantage of control-oriented reduced-order models~(ROM), similar to those described in~\citep{Kalur2019}.
Once we have computed the stabilizing SOF controller gain $F_0$ using the ROM, this can be used to initialize the Anderson-Moore algorithm (see Algorithm~\ref{Algorithm1}) for yielding an SOF-LQR control solution.

We note that the actual synthesis of SOF-LQR controllers based on Algorithm~\ref{Algorithm1} does not require further use of the ROM, as the computational demands scale with $\mathcal{O}(n^3)$  (see Appendix~\ref{sec:app2}).
Although we do not make use of a ROM for the SOF-LQR design in this study, the robustness properties of SOF-LQR control make ROM-based design a potential alternative to designs based on the full-order model.
Using a ROM can allow for additional speed-up in the SOF-LQR design, if it is required. 
\section{Computational complexity of the accelerated Anderson-Moore algorithm} \label{sec:app2}
For one iteration, the complexity of the Anderson-Moore algorithm with Armijo-type adaptation is of $\mathcal{O}(n^3)$. The computational complexity of the dominant operators are listed in Table~\ref{tab:computational}, where $n$ and $p$ are the dimensions of state and output vectors, respectively. Recall that $S(F)$ and $H(F)$ are solutions of~\eqnref{eqn:solve_for_S} and~\eqnref{eqn:solve_for_H}. 

\begin{table}
\caption{Computational complexity of Algorithm 1}
\begin{tabular}{ |p{6.5cm}|p{5.5cm}|p{2.5cm}| }
\hline
\textbf{ Calculation step} & \textbf{Dominant operator} & \textbf{Complexity} \\
\hline
Solve for $S(F)$ as a function of $F$ &  Lyapunov equation~\eqnref{eqn:solve_for_S} & $n^3$ \\\hline
Solve for $H(F)$ as a function of $F$ &  Lyapunov equation~\eqnref{eqn:solve_for_H} & $n^3$ \\\hline
Determine descent direction & Matrix inverse & $p^3$ \\
\hline
Objective function evaluation& Trace of matrix multiplication & $n^3$ \\ \hline
Derivative of objective function & Two Lyapunov equations & $2n^3$ \\
\hline
\end{tabular} \label{tab:computational}
\end{table}

\bibliography{soflqr}

\begin{thebibliography}{37}%
\makeatletter
\providecommand \@ifxundefined [1]{%
 \@ifx{#1\undefined}
}%
\providecommand \@ifnum [1]{%
 \ifnum #1\expandafter \@firstoftwo
 \else \expandafter \@secondoftwo
 \fi
}%
\providecommand \@ifx [1]{%
 \ifx #1\expandafter \@firstoftwo
 \else \expandafter \@secondoftwo
 \fi
}%
\providecommand \natexlab [1]{#1}%
\providecommand \enquote  [1]{``#1''}%
\providecommand \bibnamefont  [1]{#1}%
\providecommand \bibfnamefont [1]{#1}%
\providecommand \citenamefont [1]{#1}%
\providecommand \href@noop [0]{\@secondoftwo}%
\providecommand \href [0]{\begingroup \@sanitize@url \@href}%
\providecommand \@href[1]{\@@startlink{#1}\@@href}%
\providecommand \@@href[1]{\endgroup#1\@@endlink}%
\providecommand \@sanitize@url [0]{\catcode `\\12\catcode `\$12\catcode
  `\&12\catcode `\#12\catcode `\^12\catcode `\_12\catcode `\%12\relax}%
\providecommand \@@startlink[1]{}%
\providecommand \@@endlink[0]{}%
\providecommand \url  [0]{\begingroup\@sanitize@url \@url }%
\providecommand \@url [1]{\endgroup\@href {#1}{\urlprefix }}%
\providecommand \urlprefix  [0]{URL }%
\providecommand \Eprint [0]{\href }%
\providecommand \doibase [0]{https://doi.org/}%
\providecommand \selectlanguage [0]{\@gobble}%
\providecommand \bibinfo  [0]{\@secondoftwo}%
\providecommand \bibfield  [0]{\@secondoftwo}%
\providecommand \translation [1]{[#1]}%
\providecommand \BibitemOpen [0]{}%
\providecommand \bibitemStop [0]{}%
\providecommand \bibitemNoStop [0]{.\EOS\space}%
\providecommand \EOS [0]{\spacefactor3000\relax}%
\providecommand \BibitemShut  [1]{\csname bibitem#1\endcsname}%
\let\auto@bib@innerbib\@empty
\bibitem [{\citenamefont {Schmid}\ and\ \citenamefont
  {Henningson}(2001)}]{Schmid2001}%
  \BibitemOpen
  \bibfield  {author} {\bibinfo {author} {\bibfnamefont {P.~J.}\ \bibnamefont
  {Schmid}}\ and\ \bibinfo {author} {\bibfnamefont {D.~S.}\ \bibnamefont
  {Henningson}},\ }\href@noop {} {\emph {\bibinfo {title} {Stability and
  Transition in Shear Flows}}}\ (\bibinfo  {publisher} {Springer-Verlag},\
  \bibinfo {address} {New York},\ \bibinfo {year} {2001})\BibitemShut {NoStop}%
\bibitem [{\citenamefont {Schmid}(2007)}]{Schmid2007}%
  \BibitemOpen
  \bibfield  {author} {\bibinfo {author} {\bibfnamefont {P.~J.}\ \bibnamefont
  {Schmid}},\ }\bibfield  {title} {\bibinfo {title} {Nonmodal stability
  theory},\ }\href@noop {} {\bibfield  {journal} {\bibinfo  {journal} {Annual
  Review of Fluid Mechanics}\ }\textbf {\bibinfo {volume} {39}},\ \bibinfo
  {pages} {129} (\bibinfo {year} {2007})}\BibitemShut {NoStop}%
\bibitem [{\citenamefont {Trefethen}\ \emph {et~al.}(1993)\citenamefont
  {Trefethen}, \citenamefont {Trefethen}, \citenamefont {Reddy},\ and\
  \citenamefont {Driscoll}}]{Trefethen1993}%
  \BibitemOpen
  \bibfield  {author} {\bibinfo {author} {\bibfnamefont {L.~N.}\ \bibnamefont
  {Trefethen}}, \bibinfo {author} {\bibfnamefont {A.~E.}\ \bibnamefont
  {Trefethen}}, \bibinfo {author} {\bibfnamefont {S.~C.}\ \bibnamefont
  {Reddy}},\ and\ \bibinfo {author} {\bibfnamefont {T.~A.}\ \bibnamefont
  {Driscoll}},\ }\bibfield  {title} {\bibinfo {title} {{Hydrodynamic Stability
  Without Eigenvalues}},\ }\href {https://doi.org/10.1126/science.261.5121.578}
  {\bibfield  {journal} {\bibinfo  {journal} {Science}\ }\textbf {\bibinfo
  {volume} {261}},\ \bibinfo {pages} {578} (\bibinfo {year}
  {1993})}\BibitemShut {NoStop}%
\bibitem [{\citenamefont {Reddy}\ and\ \citenamefont
  {Henningson}(1993)}]{Reddy1993}%
  \BibitemOpen
  \bibfield  {author} {\bibinfo {author} {\bibfnamefont {S.}~\bibnamefont
  {Reddy}}\ and\ \bibinfo {author} {\bibfnamefont {D.}~\bibnamefont
  {Henningson}},\ }\bibfield  {title} {\bibinfo {title} {Energy growth in
  viscous channel flows},\ }\href@noop {} {\bibfield  {journal} {\bibinfo
  {journal} {J. Fluid Mechanics}\ }\textbf {\bibinfo {volume} {252}},\ \bibinfo
  {pages} {209} (\bibinfo {year} {1993})}\BibitemShut {NoStop}%
\bibitem [{\citenamefont {Jovanovi\'{c}}\ and\ \citenamefont
  {Bamieh}(2005)}]{jovanovic2005}%
  \BibitemOpen
  \bibfield  {author} {\bibinfo {author} {\bibfnamefont {M.~R.}\ \bibnamefont
  {Jovanovi\'{c}}}\ and\ \bibinfo {author} {\bibfnamefont {B.}~\bibnamefont
  {Bamieh}},\ }\bibfield  {title} {\bibinfo {title} {Componentwise energy
  amplification in channel flows},\ }\href@noop {} {\bibfield  {journal}
  {\bibinfo  {journal} {Journal of Fluid Mechanics}\ }\textbf {\bibinfo
  {volume} {534}},\ \bibinfo {pages} {145–183} (\bibinfo {year}
  {2005})}\BibitemShut {NoStop}%
\bibitem [{\citenamefont {Butler}\ and\ \citenamefont
  {Farrell}(1992{\natexlab{a}})}]{Butler1992}%
  \BibitemOpen
  \bibfield  {author} {\bibinfo {author} {\bibfnamefont {K.~M.}\ \bibnamefont
  {Butler}}\ and\ \bibinfo {author} {\bibfnamefont {B.}~\bibnamefont
  {Farrell}},\ }\bibfield  {title} {\bibinfo {title} {Three-dimensional optimal
  perturbations in viscous shear flow},\ }\href@noop {} {\bibfield  {journal}
  {\bibinfo  {journal} {Physics of Fluids A: Fluid Dynamics}\ }\textbf
  {\bibinfo {volume} {4}},\ \bibinfo {pages} {1637} (\bibinfo {year}
  {1992}{\natexlab{a}})}\BibitemShut {NoStop}%
\bibitem [{\citenamefont {Bagheri}\ \emph {et~al.}(2009)\citenamefont
  {Bagheri}, \citenamefont {Henningson}, \citenamefont {Hœpffner},\ and\
  \citenamefont {Schmid}}]{Bagheri2009}%
  \BibitemOpen
  \bibfield  {author} {\bibinfo {author} {\bibfnamefont {S.}~\bibnamefont
  {Bagheri}}, \bibinfo {author} {\bibfnamefont {D.~S.}\ \bibnamefont
  {Henningson}}, \bibinfo {author} {\bibfnamefont {J.}~\bibnamefont
  {Hœpffner}},\ and\ \bibinfo {author} {\bibfnamefont {P.~J.}\ \bibnamefont
  {Schmid}},\ }\bibfield  {title} {\bibinfo {title} {Input-output analysis and
  control design applied to a linear model of spatially developing flows},\
  }\href@noop {} {\bibfield  {journal} {\bibinfo  {journal} {Applied Mechanics
  Review}\ }\textbf {\bibinfo {volume} {62}},\ \bibinfo {pages} {020803}
  (\bibinfo {year} {2009})}\BibitemShut {NoStop}%
\bibitem [{\citenamefont {Bagheri}\ and\ \citenamefont
  {Henningson}(2011)}]{bagheri2011}%
  \BibitemOpen
  \bibfield  {author} {\bibinfo {author} {\bibfnamefont {S.}~\bibnamefont
  {Bagheri}}\ and\ \bibinfo {author} {\bibfnamefont {D.~S.}\ \bibnamefont
  {Henningson}},\ }\bibfield  {title} {\bibinfo {title} {Transition delay using
  control theory},\ }\href@noop {} {\bibfield  {journal} {\bibinfo  {journal}
  {Philosophical Transactions of the Royal Society of London A: Mathematical,
  Physical and Engineering Sciences}\ }\textbf {\bibinfo {volume} {369}},\
  \bibinfo {pages} {1365} (\bibinfo {year} {2011})}\BibitemShut {NoStop}%
\bibitem [{\citenamefont {Joshi}\ \emph {et~al.}(1997)\citenamefont {Joshi},
  \citenamefont {Speyer},\ and\ \citenamefont {Kim}}]{joshi1997}%
  \BibitemOpen
  \bibfield  {author} {\bibinfo {author} {\bibfnamefont {S.~S.}\ \bibnamefont
  {Joshi}}, \bibinfo {author} {\bibfnamefont {J.~L.}\ \bibnamefont {Speyer}},\
  and\ \bibinfo {author} {\bibfnamefont {J.}~\bibnamefont {Kim}},\ }\bibfield
  {title} {\bibinfo {title} {A systems theory approach to the feedback
  stabilization of infinitesimal and finite-amplitude disturbances in plane
  poiseuille flow},\ }\href@noop {} {\bibfield  {journal} {\bibinfo  {journal}
  {Journal of Fluid Mechanics}\ }\textbf {\bibinfo {volume} {332}},\ \bibinfo
  {pages} {157–184} (\bibinfo {year} {1997})}\BibitemShut {NoStop}%
\bibitem [{\citenamefont {Bewley}\ and\ \citenamefont
  {Liu}(1998)}]{Bewley1998}%
  \BibitemOpen
  \bibfield  {author} {\bibinfo {author} {\bibfnamefont {T.~R.}\ \bibnamefont
  {Bewley}}\ and\ \bibinfo {author} {\bibfnamefont {S.}~\bibnamefont {Liu}},\
  }\bibfield  {title} {\bibinfo {title} {Optimal and robust control and
  estimation of linear paths to transition},\ }\href@noop {} {\bibfield
  {journal} {\bibinfo  {journal} {Journal of Fluid Mechanics}\ }\textbf
  {\bibinfo {volume} {365}},\ \bibinfo {pages} {305} (\bibinfo {year}
  {1998})}\BibitemShut {NoStop}%
\bibitem [{\citenamefont {H\"{o}gberg}\ \emph {et~al.}(2003)\citenamefont
  {H\"{o}gberg}, \citenamefont {Bewley},\ and\ \citenamefont
  {Henningson}}]{hogberg2003}%
  \BibitemOpen
  \bibfield  {author} {\bibinfo {author} {\bibfnamefont {M.}~\bibnamefont
  {H\"{o}gberg}}, \bibinfo {author} {\bibfnamefont {T.~R.}\ \bibnamefont
  {Bewley}},\ and\ \bibinfo {author} {\bibfnamefont {D.~S.}\ \bibnamefont
  {Henningson}},\ }\bibfield  {title} {\bibinfo {title} {Linear feedback
  control and estimation of transition in plane channel flow},\ }\href@noop {}
  {\bibfield  {journal} {\bibinfo  {journal} {Journal of Fluid Mechanics}\
  }\textbf {\bibinfo {volume} {481}},\ \bibinfo {pages} {149–175} (\bibinfo
  {year} {2003})}\BibitemShut {NoStop}%
\bibitem [{\citenamefont {Martinelli}\ \emph {et~al.}(2011)\citenamefont
  {Martinelli}, \citenamefont {Quadrio}, \citenamefont {Mc{K}ernan},\ and\
  \citenamefont {Whidborne}}]{Martinelli2011}%
  \BibitemOpen
  \bibfield  {author} {\bibinfo {author} {\bibfnamefont {F.}~\bibnamefont
  {Martinelli}}, \bibinfo {author} {\bibfnamefont {M.}~\bibnamefont {Quadrio}},
  \bibinfo {author} {\bibfnamefont {J.}~\bibnamefont {Mc{K}ernan}},\ and\
  \bibinfo {author} {\bibfnamefont {J.~F.}\ \bibnamefont {Whidborne}},\
  }\bibfield  {title} {\bibinfo {title} {Linear feedback control of transient
  energy growth and control performance limitations in subcritical plane
  poiseuille flow},\ }\href@noop {} {\bibfield  {journal} {\bibinfo  {journal}
  {Physics of Fluids}\ }\textbf {\bibinfo {volume} {23}},\ \bibinfo {pages}
  {014103} (\bibinfo {year} {2011})}\BibitemShut {NoStop}%
\bibitem [{\citenamefont {Sun}\ and\ \citenamefont {Hemati}(2019)}]{sun2019}%
  \BibitemOpen
  \bibfield  {author} {\bibinfo {author} {\bibfnamefont {Y.}~\bibnamefont
  {Sun}}\ and\ \bibinfo {author} {\bibfnamefont {M.~S.}\ \bibnamefont
  {Hemati}},\ }\bibfield  {title} {\bibinfo {title} {Feedback control for
  transition suppression in direct numerical simulations of channel flow},\
  }\href@noop {} {\bibfield  {journal} {\bibinfo  {journal} {Energies}\
  }\textbf {\bibinfo {volume} {12}} (\bibinfo {year} {2019})}\BibitemShut
  {NoStop}%
\bibitem [{\citenamefont {Ilak}\ and\ \citenamefont {Rowley}(2008)}]{Ilak2008}%
  \BibitemOpen
  \bibfield  {author} {\bibinfo {author} {\bibfnamefont {M.}~\bibnamefont
  {Ilak}}\ and\ \bibinfo {author} {\bibfnamefont {C.~W.}\ \bibnamefont
  {Rowley}},\ }\bibfield  {title} {\bibinfo {title} {Feedback control of
  transitional channel flow using balanced proper orthogonal decomposition},\
  }\href@noop {} {\bibfield  {journal} {\bibinfo  {journal} {AIAA paper
  2008-4230}\ } (\bibinfo {year} {2008})}\BibitemShut {NoStop}%
\bibitem [{\citenamefont {Kalur}\ and\ \citenamefont
  {Hemati}(2019)}]{Kalur2019}%
  \BibitemOpen
  \bibfield  {author} {\bibinfo {author} {\bibfnamefont {A.}~\bibnamefont
  {Kalur}}\ and\ \bibinfo {author} {\bibfnamefont {M.~S.}\ \bibnamefont
  {Hemati}},\ }\bibfield  {title} {\bibinfo {title} {Control-oriented model
  reduction for minimizing transient energy growth in shear flows},\
  }\href@noop {} {\bibfield  {journal} {\bibinfo  {journal} {AIAA Journal}\ }
  (\bibinfo {year} {2019})}\BibitemShut {NoStop}%
\bibitem [{\citenamefont {Brogan}(1991)}]{Brogan1991}%
  \BibitemOpen
  \bibfield  {author} {\bibinfo {author} {\bibfnamefont {W.~L.}\ \bibnamefont
  {Brogan}},\ }\href@noop {} {\emph {\bibinfo {title} {Modern Control
  Theory}}}\ (\bibinfo  {publisher} {Prentice Hall},\ \bibinfo {year}
  {1991})\BibitemShut {NoStop}%
\bibitem [{\citenamefont {Hemati}\ and\ \citenamefont
  {Yao}(2018)}]{Hemati2018}%
  \BibitemOpen
  \bibfield  {author} {\bibinfo {author} {\bibfnamefont {M.~S.}\ \bibnamefont
  {Hemati}}\ and\ \bibinfo {author} {\bibfnamefont {H.}~\bibnamefont {Yao}},\
  }\bibfield  {title} {\bibinfo {title} {Performance limitations of
  observer-based feedback for transient energy growth suppression},\
  }\href@noop {} {\bibfield  {journal} {\bibinfo  {journal} {AIAA Journal}\
  }\textbf {\bibinfo {volume} {56}},\ \bibinfo {pages} {2119} (\bibinfo {year}
  {2018})}\BibitemShut {NoStop}%
\bibitem [{\citenamefont {Yao}\ and\ \citenamefont {Hemati}(2018)}]{Yao2018}%
  \BibitemOpen
  \bibfield  {author} {\bibinfo {author} {\bibfnamefont {H.}~\bibnamefont
  {Yao}}\ and\ \bibinfo {author} {\bibfnamefont {M.~S.}\ \bibnamefont
  {Hemati}},\ }\bibfield  {title} {\bibinfo {title} {Revisiting the separation
  principle for improved transition control}\ }(\bibinfo {year}
  {2018})\BibitemShut {NoStop}%
\bibitem [{\citenamefont {Whidborne}\ \emph {et~al.}(2005)\citenamefont
  {Whidborne}, \citenamefont {Mc{K}ernan},\ and\ \citenamefont
  {Steer}}]{Whidborne2005}%
  \BibitemOpen
  \bibfield  {author} {\bibinfo {author} {\bibfnamefont {J.~F.}\ \bibnamefont
  {Whidborne}}, \bibinfo {author} {\bibfnamefont {J.}~\bibnamefont
  {Mc{K}ernan}},\ and\ \bibinfo {author} {\bibfnamefont {A.~J.}\ \bibnamefont
  {Steer}},\ }\bibfield  {title} {\bibinfo {title} {Minimization of maximum
  transient energy growth by output feedback},\ }\href@noop {} {\bibfield
  {journal} {\bibinfo  {journal} {IFAC Proceedings}\ }\textbf {\bibinfo
  {volume} {38}},\ \bibinfo {pages} {283} (\bibinfo {year} {2005})}\BibitemShut
  {NoStop}%
\bibitem [{\citenamefont {Whidborne}\ and\ \citenamefont
  {Mc{K}ernan}(2007)}]{Whidborne2007}%
  \BibitemOpen
  \bibfield  {author} {\bibinfo {author} {\bibfnamefont {J.~F.}\ \bibnamefont
  {Whidborne}}\ and\ \bibinfo {author} {\bibfnamefont {J.}~\bibnamefont
  {Mc{K}ernan}},\ }\bibfield  {title} {\bibinfo {title} {On the minimization of
  maximum transient energy growth},\ }\href@noop {} {\bibfield  {journal}
  {\bibinfo  {journal} {IEEE Transactions on Automatic Control}\ }\textbf
  {\bibinfo {volume} {52}},\ \bibinfo {pages} {1762} (\bibinfo {year}
  {2007})}\BibitemShut {NoStop}%
\bibitem [{\citenamefont {Hœpffner}\ \emph {et~al.}(2005)\citenamefont
  {Hœpffner}, \citenamefont {Chevalier}, \citenamefont {Bewley},\ and\
  \citenamefont {Henningson}}]{hapffner2005}%
  \BibitemOpen
  \bibfield  {author} {\bibinfo {author} {\bibfnamefont {J.}~\bibnamefont
  {Hœpffner}}, \bibinfo {author} {\bibfnamefont {M.}~\bibnamefont
  {Chevalier}}, \bibinfo {author} {\bibfnamefont {T.~R.}\ \bibnamefont
  {Bewley}},\ and\ \bibinfo {author} {\bibfnamefont {D.~S.}\ \bibnamefont
  {Henningson}},\ }\bibfield  {title} {\bibinfo {title} {State estimation in
  wall-bounded flow systems. part 1. perturbed laminar flows},\ }\href@noop {}
  {\bibfield  {journal} {\bibinfo  {journal} {Journal of Fluid Mechanics}\
  }\textbf {\bibinfo {volume} {534}},\ \bibinfo {pages} {263–294} (\bibinfo
  {year} {2005})}\BibitemShut {NoStop}%
\bibitem [{\citenamefont {Anderson}\ and\ \citenamefont
  {Moore}(1971)}]{Anderson1971}%
  \BibitemOpen
  \bibfield  {author} {\bibinfo {author} {\bibfnamefont {B.~D.}\ \bibnamefont
  {Anderson}}\ and\ \bibinfo {author} {\bibfnamefont {J.~B.}\ \bibnamefont
  {Moore}},\ }\href@noop {} {\emph {\bibinfo {title} {Linear Optimal
  Control}}}\ (\bibinfo  {publisher} {Prentice-Hall},\ \bibinfo {address} {New
  Jersey},\ \bibinfo {year} {1971})\BibitemShut {NoStop}%
\bibitem [{\citenamefont {Butler}\ and\ \citenamefont
  {Farrell}(1992{\natexlab{b}})}]{butlerPOF1992}%
  \BibitemOpen
  \bibfield  {author} {\bibinfo {author} {\bibfnamefont {K.~M.}\ \bibnamefont
  {Butler}}\ and\ \bibinfo {author} {\bibfnamefont {B.~F.}\ \bibnamefont
  {Farrell}},\ }\bibfield  {title} {\bibinfo {title} {Three‐dimensional
  optimal perturbations in viscous shear flow},\ }\href@noop {} {\bibfield
  {journal} {\bibinfo  {journal} {Physics of Fluids A: Fluid Dynamics}\
  }\textbf {\bibinfo {volume} {4}},\ \bibinfo {pages} {1637} (\bibinfo {year}
  {1992}{\natexlab{b}})}\BibitemShut {NoStop}%
\bibitem [{\citenamefont {Yao}\ and\ \citenamefont {Hemati}(2019)}]{Yao2019}%
  \BibitemOpen
  \bibfield  {author} {\bibinfo {author} {\bibfnamefont {H.}~\bibnamefont
  {Yao}}\ and\ \bibinfo {author} {\bibfnamefont {M.~S.}\ \bibnamefont
  {Hemati}},\ }\bibfield  {title} {\bibinfo {title} {Advances in output
  feedback control of transient energy growth in a linearized channel flow},\
  }\href@noop {} {\bibfield  {journal} {\bibinfo  {journal} {AIAA Paper
  2019-0882}\ } (\bibinfo {year} {2019})}\BibitemShut {NoStop}%
\bibitem [{\citenamefont {Choi}\ \emph {et~al.}(1994)\citenamefont {Choi},
  \citenamefont {Moin},\ and\ \citenamefont {Kim}}]{choi1994}%
  \BibitemOpen
  \bibfield  {author} {\bibinfo {author} {\bibfnamefont {H.}~\bibnamefont
  {Choi}}, \bibinfo {author} {\bibfnamefont {P.}~\bibnamefont {Moin}},\ and\
  \bibinfo {author} {\bibfnamefont {J.}~\bibnamefont {Kim}},\ }\bibfield
  {title} {\bibinfo {title} {Active turbulence control for drag reduction in
  wall-bounded flows},\ }\href@noop {} {\bibfield  {journal} {\bibinfo
  {journal} {Journal of Fluid Mechanics}\ }\textbf {\bibinfo {volume} {262}},\
  \bibinfo {pages} {75} (\bibinfo {year} {1994})}\BibitemShut {NoStop}%
\bibitem [{\citenamefont {Luhar}\ \emph {et~al.}(2014)\citenamefont {Luhar},
  \citenamefont {Sharma},\ and\ \citenamefont {McKeon}}]{luhar2014}%
  \BibitemOpen
  \bibfield  {author} {\bibinfo {author} {\bibfnamefont {M.}~\bibnamefont
  {Luhar}}, \bibinfo {author} {\bibfnamefont {A.}~\bibnamefont {Sharma}},\ and\
  \bibinfo {author} {\bibfnamefont {B.}~\bibnamefont {McKeon}},\ }\bibfield
  {title} {\bibinfo {title} {Opposition control within the resolvent analysis
  framework},\ }\href@noop {} {\bibfield  {journal} {\bibinfo  {journal}
  {Journal of Fluid Mechanics}\ }\textbf {\bibinfo {volume} {749}},\ \bibinfo
  {pages} {597} (\bibinfo {year} {2014})}\BibitemShut {NoStop}%
\bibitem [{\citenamefont {Rautert}\ and\ \citenamefont
  {Sachs}(1997)}]{Rautert1997}%
  \BibitemOpen
  \bibfield  {author} {\bibinfo {author} {\bibfnamefont {T.}~\bibnamefont
  {Rautert}}\ and\ \bibinfo {author} {\bibfnamefont {E.~W.}\ \bibnamefont
  {Sachs}},\ }\bibfield  {title} {\bibinfo {title} {{Computational Design of
  Optimal Output Feedback Controllers}},\ }\href@noop {} {\bibfield  {journal}
  {\bibinfo  {journal} {SIAM J. Optimization}\ }\textbf {\bibinfo {volume}
  {7}},\ \bibinfo {pages} {837} (\bibinfo {year} {1997})}\BibitemShut {NoStop}%
\bibitem [{\citenamefont {Syrmos}\ \emph {et~al.}(1997)\citenamefont {Syrmos},
  \citenamefont {Abdallah}, \citenamefont {Dorato},\ and\ \citenamefont
  {Grigoriadis}}]{Syrmos1997}%
  \BibitemOpen
  \bibfield  {author} {\bibinfo {author} {\bibfnamefont {V.~L.}\ \bibnamefont
  {Syrmos}}, \bibinfo {author} {\bibfnamefont {C.~T.}\ \bibnamefont
  {Abdallah}}, \bibinfo {author} {\bibfnamefont {P.}~\bibnamefont {Dorato}},\
  and\ \bibinfo {author} {\bibfnamefont {K.}~\bibnamefont {Grigoriadis}},\
  }\bibfield  {title} {\bibinfo {title} {{Static Output Feedback: A Survey}},\
  }\href {https://doi.org/10.1016/S0005-1098(96)00141-0} {\bibfield  {journal}
  {\bibinfo  {journal} {Automatica}\ }\textbf {\bibinfo {volume} {33}},\
  \bibinfo {pages} {125} (\bibinfo {year} {1997})}\BibitemShut {NoStop}%
\bibitem [{\citenamefont {Nocedal}\ and\ \citenamefont
  {Wright}(2006)}]{nocedal2006}%
  \BibitemOpen
  \bibfield  {author} {\bibinfo {author} {\bibfnamefont {J.}~\bibnamefont
  {Nocedal}}\ and\ \bibinfo {author} {\bibfnamefont {S.~J.}\ \bibnamefont
  {Wright}},\ }\href@noop {} {\emph {\bibinfo {title} {Numerical
  Optimization}}},\ \bibinfo {edition} {2nd}\ ed.\ (\bibinfo  {publisher}
  {Springer},\ \bibinfo {address} {New York},\ \bibinfo {year}
  {2006})\BibitemShut {NoStop}%
\bibitem [{\citenamefont {Toivonen}\ and\ \citenamefont
  {M{\"{a}}kil{\"{a}}}(1985)}]{Toivonen1985}%
  \BibitemOpen
  \bibfield  {author} {\bibinfo {author} {\bibfnamefont {H.~T.}\ \bibnamefont
  {Toivonen}}\ and\ \bibinfo {author} {\bibfnamefont {P.~M.}\ \bibnamefont
  {M{\"{a}}kil{\"{a}}}},\ }\bibfield  {title} {\bibinfo {title} {{A Descent
  Anderson-Moore Algorithm for Optimal Decentralized Control}},\ }\href@noop {}
  {\bibfield  {journal} {\bibinfo  {journal} {Automatica}\ }\textbf {\bibinfo
  {volume} {21}},\ \bibinfo {pages} {743} (\bibinfo {year} {1985})}\BibitemShut
  {NoStop}%
\bibitem [{\citenamefont {Cao}\ \emph {et~al.}(1998)\citenamefont {Cao},
  \citenamefont {Lam},\ and\ \citenamefont {Sun}}]{Cao1998a}%
  \BibitemOpen
  \bibfield  {author} {\bibinfo {author} {\bibfnamefont {Y.-Y.}\ \bibnamefont
  {Cao}}, \bibinfo {author} {\bibfnamefont {J.}~\bibnamefont {Lam}},\ and\
  \bibinfo {author} {\bibfnamefont {Y.-X.}\ \bibnamefont {Sun}},\ }\bibfield
  {title} {\bibinfo {title} {{Static Output Feedback Stabilization: An ILMI
  Approach}},\ }\href {https://doi.org/10.1016/S0005-1098(98)80021-6}
  {\bibfield  {journal} {\bibinfo  {journal} {Automatica}\ }\textbf {\bibinfo
  {volume} {34}},\ \bibinfo {pages} {1641} (\bibinfo {year}
  {1998})}\BibitemShut {NoStop}%
\bibitem [{\citenamefont {Mc{K}ernan}\ \emph {et~al.}(2006)\citenamefont
  {Mc{K}ernan}, \citenamefont {Whidborne},\ and\ \citenamefont
  {Papadakis}}]{McKernanJ2006}%
  \BibitemOpen
  \bibfield  {author} {\bibinfo {author} {\bibfnamefont {J.}~\bibnamefont
  {Mc{K}ernan}}, \bibinfo {author} {\bibfnamefont {J.}~\bibnamefont
  {Whidborne}},\ and\ \bibinfo {author} {\bibfnamefont {G.}~\bibnamefont
  {Papadakis}},\ }\bibfield  {title} {\bibinfo {title} {A linear state-space
  representation of plane poiseuille flow for control design -- a tutorial},\
  }\href@noop {} {\bibfield  {journal} {\bibinfo  {journal} {International
  Journal of Modelling Identification and Control}\ }\textbf {\bibinfo {volume}
  {1}},\ \bibinfo {pages} {272} (\bibinfo {year} {2006})}\BibitemShut {NoStop}%
\bibitem [{\citenamefont {Gibson}(2014)}]{channelflow}%
  \BibitemOpen
  \bibfield  {author} {\bibinfo {author} {\bibfnamefont {J.~F.}\ \bibnamefont
  {Gibson}},\ }\href@noop {} {\emph {\bibinfo {title} {{Channelflow}: {A}
  spectral {Navier-Stokes} simulator in {C}++}}},\ \bibinfo {type} {Tech.
  Rep.}\ (\bibinfo  {institution} {U. New Hampshire},\ \bibinfo {year} {2014})\
  \bibinfo {note} {{\tt {Channelflow.org}}}\BibitemShut {NoStop}%
\bibitem [{\citenamefont {Reddy}\ \emph {et~al.}(1998)\citenamefont {Reddy},
  \citenamefont {Schmid}, \citenamefont {baggett},\ and\ \citenamefont
  {Henningson}}]{Reddy:JFM98}%
  \BibitemOpen
  \bibfield  {author} {\bibinfo {author} {\bibfnamefont {S.~C.}\ \bibnamefont
  {Reddy}}, \bibinfo {author} {\bibfnamefont {P.~J.}\ \bibnamefont {Schmid}},
  \bibinfo {author} {\bibfnamefont {J.~S.}\ \bibnamefont {baggett}},\ and\
  \bibinfo {author} {\bibfnamefont {D.~S.}\ \bibnamefont {Henningson}},\
  }\bibfield  {title} {\bibinfo {title} {On stability of streamwise streaks and
  transition thresholds in plane channel flows},\ }\href@noop {} {\bibfield
  {journal} {\bibinfo  {journal} {Journal of Fluid Mechanics}\ }\textbf
  {\bibinfo {volume} {365}},\ \bibinfo {pages} {269} (\bibinfo {year}
  {1998})}\BibitemShut {NoStop}%
\bibitem [{\citenamefont {Orszag}\ and\ \citenamefont
  {Kells}(1980)}]{orszag_kells_1980}%
  \BibitemOpen
  \bibfield  {author} {\bibinfo {author} {\bibfnamefont {S.~A.}\ \bibnamefont
  {Orszag}}\ and\ \bibinfo {author} {\bibfnamefont {L.~C.}\ \bibnamefont
  {Kells}},\ }\bibfield  {title} {\bibinfo {title} {Transition to turbulence in
  plane poiseuille and plane couette flow},\ }\href
  {https://doi.org/10.1017/S0022112080002066} {\bibfield  {journal} {\bibinfo
  {journal} {Journal of Fluid Mechanics}\ }\textbf {\bibinfo {volume} {96}},\
  \bibinfo {pages} {159–205} (\bibinfo {year} {1980})}\BibitemShut {NoStop}%
\bibitem [{\citenamefont {Orszag}(1971)}]{orszag_1971}%
  \BibitemOpen
  \bibfield  {author} {\bibinfo {author} {\bibfnamefont {S.~A.}\ \bibnamefont
  {Orszag}},\ }\bibfield  {title} {\bibinfo {title} {Accurate solution of the
  orr–sommerfeld stability equation},\ }\href
  {https://doi.org/10.1017/S0022112071002842} {\bibfield  {journal} {\bibinfo
  {journal} {Journal of Fluid Mechanics}\ }\textbf {\bibinfo {volume} {50}},\
  \bibinfo {pages} {689–703} (\bibinfo {year} {1971})}\BibitemShut {NoStop}%
\bibitem [{\citenamefont {Hunt}\ \emph {et~al.}(1988)\citenamefont {Hunt},
  \citenamefont {Wray},\ and\ \citenamefont {Moin}}]{Hunt:88}%
  \BibitemOpen
  \bibfield  {author} {\bibinfo {author} {\bibfnamefont {J.~C.~R.}\
  \bibnamefont {Hunt}}, \bibinfo {author} {\bibfnamefont {A.~A.}\ \bibnamefont
  {Wray}},\ and\ \bibinfo {author} {\bibfnamefont {P.}~\bibnamefont {Moin}},\
  }\bibfield  {title} {\bibinfo {title} {Eddies, streams, and convergence zones
  in turbulent flows},\ }\href@noop {} {\bibfield  {journal} {\bibinfo
  {journal} {Proc. of the Summer Program, Center of Turbulence Research}\ ,\
  \bibinfo {pages} {193}} (\bibinfo {year} {1988})}\BibitemShut {NoStop}%
\end{thebibliography}%

\end{document}